\documentclass[preprint,aps,showpacs]{revtex4}
\usepackage{graphicx}
\usepackage{mathrsfs} 
\usepackage{float}
\usepackage{subfigure}
\usepackage{colordvi}
\usepackage{amssymb}
\usepackage{amsmath}
\usepackage{bbold}
\usepackage{epsf}
\begin{document}
\def \beq{\begin{equation}}
\def \eeq{\end{equation}}
\def \bea{\begin{eqnarray}}
\def \eea{\end{eqnarray}}
\def \bem{\begin{displaymath}}
\def \eem{\end{displaymath}}
\def \P{\Psi}
\def \Pd{|\Psi(\boldsymbol{r})|}
\def \Pds{|\Psi^{\ast}(\boldsymbol{r})|}
\def \Po{\overline{\Psi}}
\def \bs{\boldsymbol}
\def \bl{\bar{\boldsymbol{l}}}

\title{Electron optics with dirac fermions: electron transport in monolayer and bilayer graphene
through magnetic barrier and their superlattices} 
\author{Neetu Agrawal (Garg)$^1$, Sankalpa  Ghosh$^2$ and  Manish Sharma$^1$}
\affiliation{$^1$Centre for Applied Research in Electronics, Indian Institute of Technology Delhi, New Delhi-110016, India}
\affiliation{$^2$Department of Physics, Indian Institute of Technology Delhi, New Delhi-110016, India}
\email {sankalpa@physics.iitd.ac.in}
\begin{abstract}
In this  review article we discuss the recent progress in studying ballistic transport for charge carriers in graphene through 
highly inhomogenous magnetic field known as magnetic barrier in combination with gate voltage induced electrostatic potential. Starting with  cases for a single or double magnetic barrier we also review the progress in understanding 
electron transport through the superlattices created out of such electromagnetic potential barriers and discuss the 
possibility of experimental realization of such systems. 
The emphasis is particularly on the analogy of such transport 
with propagation of light wave through medium with alternating dielectric constant. In that direction 
we discuss electron analogue of optical phenomena like fabry perot resonances, 
negative refraction, Goos-H\"anchen effect, beam collimation in such systems and explain how such analogy is going to be useful for device generation. The resulting modification of band structure of dirac fermions, the emergence of additional 
dirac points was also discussed accompanied by brief section on the interconvertibility of electric and magnetic field for relativistic dirac fermions. We also discuss the effect of such electromagnetic potential barrier on bilayer graphene in a similar framework. 
\end{abstract}
\pacs{ 81.05. Ue, 73.63.-b, 78.20.Ci, 42.25.Gy, 73.43.Qt}{}
\date{\today}
\maketitle
\tableofcontents
\section{Introduction}

Physically Maxwell's equation that describes the propagation of electromagnetic wave/light through free space and dielectric medium, namely  
\beq \frac{\partial^2 \bs{(E,B)}}{\partial t^2}  = \frac{1}{\mu \epsilon} \nabla^2 \bs{(E,B)} \label{Max} \eeq 
and Schr$\ddot{o}$dinger equation that describes the time evolution of probability amplitude or de Broglie wave in quantum systems, namely
\beq i \hbar \frac{\partial \Psi}{\partial t} = -\frac{\hbar^2}{2m} \nabla^{2} \Psi + V \Psi \label{Sch} \eeq 
correspond to the different type of phenomena. However, their striking mathematical similarity indicates that a large  number of wave like phenomena will occur in either of these cases. Whereas the motion of a isolated single electron in presence 
of potential barrier is described by the wave  Eq.(\ref{Sch}), 
in real materials due to the scattering by other electrons, impurities etc. the motion of the charge carriers is generally 
diffusive. Therefore, such wave propagation based description of the electron transport becomes only meaningful if the electron mean free path is of the order the typical  sample size. Such a transport regime is called the 
ballistic transport regime and  in that regime the similarity between the propagations of transport electrons and electromagnetic wave or light promises rich dividend \cite{dutta, analogy}.

 A particular consequence of the similarity
between the  wave equations  (\ref{Max}) and (\ref{Sch}) is that for a 
monochromatic wave with a given frequency in Eq. \ref{Max}
 the role played by the dielectric constant  of a given medium has its corresponding analogue in the the potential landscape for a stationary solution of the Schr$\ddot{o}$dinger electron described by Eq. \ref{Sch}. Thus some of the effects that one obtains by spatially 
modulating the potential in a Schr$\ddot{o}$dinger equation can be reproduced for the light also by spatially modulating the dielectric
constant of the medium. Indeed this was pointed out in seventies in the pioneering work by Yariv and collaborators 
\cite {yariv} that the propagation of light through a medium with periodically modulated dielectric constant will lead 
to the similar band structure of transport electron as observed in Kronig-Penny model \cite{Kittel}. In this context 
it is particularly useful to point out 
that for conventional  two dimensional electron gas (2DEG),  the analogy between  transmission of de Broglie waves satisfying the Schr$\ddot{o}$dinger equation through a one-dimensional electrostatic potential  \cite{SIVAN90, Stormer} and light propagation in  linear dilelectric medium is well established and led to the development of a number of applications.

The recent discovery of graphene \cite{Wallace, KSN1, KSN2, YZ1} added a new twist to this 
well established optical analogy of ballistic electron transport and vice versa. In graphene, electrons near the Fermi level, namely the transport electrons no more obeys 
quadratic dispersion law, a typical characteristics of their non-relativistic nature, but rather  obeys a 
linear dispersion relation at or near the Fermi surface, an archetypical of the ultra relativistic massless particles.
Hence they are called massless Dirac fermions. It has now been established that transport of such massless Dirac fermions 
in the presence of an electrostatic potential barrier is analogous
to negative refraction through metamaterials \cite{cheinov, Veselago, pend85}. The
relativistic behaviour of graphene electrons also leads to Klein
tunnelling \cite{KSN3}, where a  massless relativistic particle can tunnel through
any potential barrier above the Fermi surface, invalidating the
possibility of confining it using such potential barriers.

In the context of  electronic transport  in graphene devices that are proposed for graphene based electronic in future,  
 the local carrier concentration is controlled by one or more local gates. 
 Particularly,  in the cleanest devices which satisfy the criterion of ballistic transport, transport signatures
of the relativistic nature of the charge carriers of graphene are observed. A vast body of theoretical work ( for a review see
\cite{Youngreview}) has already been devoted to the study of multigated graphene superlattices which can form a basis for a new kind of electronic optics based on graphene as an electronic metamaterial. However the inability of confining the electrons using a electrostatic potential barrier severely limits the applicability of such devices. A solution to that problem 
was proposed by De Martino et al. \cite{eggerprl} when it was pointed out that well localized magnetic field dubbed as
magnetic barrier can confine massless dirac fermions in graphene. This suggests that one way of making  high mobility graphene based electronic devices is to make locally gated structures in conjunction with such magnetic barriers. 
Given this context it is extremely important to see what will happen to the above mentioned optical analogy when it applies to the cases of such magnetic barriers. 

However it may be pointed out that 
the study of transport properties of two dimensional non relativistic electron  system in the presence of a transverse 
magnetic field remains mostly confined to the case where the magnetic field is uniform on the scale of sample size through phenomena  
such as Integer and Fractional Quantum Hall effect. Study of electron transport in inhomogenous magnetic field 
is relatively lesser known field \cite{nogaretjpcm}, even though inhomogenous magnetic field is being used for many other purpose for a long time, such as one in the  famous Stern Gerlach experiment \cite{sterngerlach}.
In a homogenous magnetic field, the electrons execute cyclotron motion, the direction of the wavevector continuously changes, hence rendering any analogy with the monochromatic light propagation in dielectric medium is very unlikely.
In quantum mechanical language the electronic wavefunction is localized over the scale of the magnetic length. Thus a direct
analogy with a propagating wave is not possible.

This situation however changed, when  it was pointed out \cite{sgms} that an optical analogy can be constructed 
for electron transport in graphene in presence of highly inhomogenous magnetic field \cite{peetersprl}
where one does not get bound state solutions, but rather scattering state solutions. This  suggests that inhomogenous magnetic field not only 
confine massless dirac fermions \cite{eggerprl}, but the ballistic transport through such barriers can also  be understood in terms of 
suitable optical analogy.
Ref. \cite{sgms} also showed that the analogy with light propagation in  medium with periodically modulated dielectric constant can again be constructed by considering the transport of massless dirac fermions in graphene through the periodic 
arrangement of such highly inhomogenous magnetic field. Subsequent work \cite{mssg} showed  that a very rich transport regime 
can be realized for such graphene electrons by considering structures where a local gate voltage is used simultaneously with 
such magnetic barrier. When such configurations are generalized to a superlattice structure, that significantly 
modifies the band structure of graphene electrons. Subsequently a large volume of theoretical work pointed out the 
relevance of such optical analogy and proposed device structures based on that. Other properties such as Goos H\"anchen shift of graphene electrons by these type of barriers, collimation of the electron beam, generation of additional Dirac points 
was also discussed in further works. It was also extended in the case of bilayer graphene. 
In the current review article we plan to review these developments. The purpose 
is two fold. One, is to review the general theoretical framework in a pedagogical way, highlighting a number of  significant work done 
in this subfield of graphene research. The other purpose is to indicate in very clear terms the possibility of designing interesting experiment and 
novel device structure that holds rich promises for graphene based electronics.

The organization of the article is as follows. After giving a brief review of low energy description of charge carriers in graphene, first we explain how electron transport through scalar electrostatic barriers and magnetic barriers is explained in the language of geometrical optics.  Next, we explain how such transport gets modified when various types of potential barriers are created atop magnetic barriers. We discuss Goos-H\"anchen shift and  fabry perot resonances in this type of structure. Then we provide an analysis of transport through an infinite series of such barriers pointing towards an effective way of changing the bandstructure. Here we also discuss 
the collimation of the electron beam, emergence of extra dirac point etc. We then briefly discuss the issue of interconvertibilty of electrostatic potential and vector potential due to magnetic field citing relativistic invariance of the equation obeyed by transport electrons in graphene. 
Then we come to the issue of electron transport through graphene bilayer in presence of such potential barriers. The dispersion relation for graphene bilayer is very different from that of monolayer graphene. The charge carriers in bilayer graphene have a parabolic energy spectrum, which means they are massive, similar to conventional non relativistic electrons. On the other hand, due to the crystal structure of graphene consisting of two sublattices, these particles are described by spinor wave functions similar to that of monolayer graphene. We study how the transport of such electrons in graphene bilayer is affected, again when exposed to combination of magnetic barriers and voltages. 

\subsection{Low energy description of charge carriers in graphene} \label{general}
The interesting band structure of graphene was pointed out by Phil Wallace \cite{Wallace} way back in 1947. After its 
experimental discovery by Geim, Novoselov, Kim  and their collaborators
there are many excellent review articles (for example see \cite{Castroneto} ) where 
the band structure, the resulting dirac fermion nature of the charge carriers are detailed. Here we provide a brief introduction to the low energy description of these charge carriers for completeness. 
Graphene is a one atom thick planar sheet of $sp^2$ bonded carbon atoms. The carbon atoms are distributed at the edges of regular hexagons forming a honeycomb lattice. However, the honeycomb lattice is not a Bravais lattice and, from a crystallographic point of view, has to be described by two inter penetrating triangular lattices with two atoms (sublattice A and B) per unit cell.  

Each carbon atom has four valence electrons, three of which hybridize to form  $  sp^{2} $ hybridized orbitals to form sigma bonds with neighbouring atoms. However, the fourth unbound electron lies in the $2p_z$ orbital which extends vertically above and below the plane. It is this electron which interacts with the periodic field of the hexagonal crystal lattice of graphene. If we denote the creation operator by $a^{\dag}_i(b^{\dag}_i)$ for an atom on the A(B) sublattice, then the nearest neighbour tightbinding Hamiltonian has the simple form \cite{pereirarev,fuchs}
\beq \mathcal{H} = -t_1\sum_{i,j}(a^{\dag}_ib_j+H.c. )\eeq
where  $t_1 \approx 2.8 meV$ is the nearest neighbour hopping parameter.
The tightbinding eigen functions has the form of bispinor, whose components corresponds to the amplitudes on sublattice A and B respectively within the unit cell. This also means that in addition to its usual spin $1/2$, electrons carriers an additional pseudo-spin $1/2$ associated with its sublattice degree of freedom. The energy levels of the lattice can be determined by operating with the Hamiltonian on its eigen function to give the corresponding band structure:

\beq E = \pm t_1\sqrt{3+f(\bs{k})} \label{bandstructure} \eeq
where $f(\bs{k}) = 2\cos(q_ya\sqrt{3}) + 4 cos (q_x3a/2)\cos (q_ya\sqrt{3}/2)$ with lattice parameter, $a \approx 2.46 A^{0}$.

\begin{figure}
\begin{center}
\includegraphics[width=120 mm]{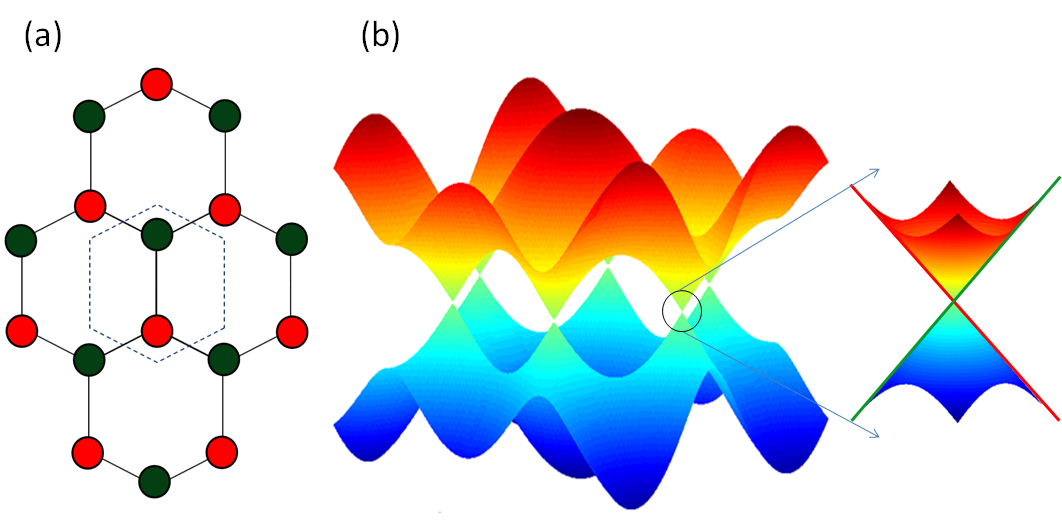}
\end{center}
\caption{(a) Honey comb lattice of graphene showing two inequivalent carbon atoms (with different colour) in a unit cell  
(b) Band structure in honey comb lattice, with zoom in of the energy bands close to one of the Dirac points.}
\label{bandcone}
\end{figure}

In Eq.(\ref{bandstructure}) the plus sign refers to the upper ($\pi^{*}$) band and the minus sign refers to the lower ($\pi$) band.
The unique feature of the band structure of graphene \cite{saito, dressdress} is that its valence band $(\pi)$ and conduction band $(\pi^{*})$ touch each other at 6 specific points, which are also the vertices of the hexagonal Brillouin zone of this honeycomb lattice ( see Fig. \ref{bandcone}).
 Out of these six points, only two $K$ and $K^{\prime}$ are non-equivalent and all the rest are related to these two points by symmetry. These points are also known as the $\bs{K}$ points in the literature and their positions in the reciprocal space are:

\bea \bs{K} &=& \left(\frac{2\pi}{3a}, \frac{2\pi}{3\sqrt{3}a}\right), \nonumber \\
\bs{ K^{\prime}} &=& \left(\frac{2\pi}{3a}, -\frac{2\pi}{3\sqrt{3}a}\right) \nonumber \eea

Under ambient conditions, also known in the literature as zero biased doping, the Fermi level coincide with these corners of the hexagonal Brillouin zone. Thus, to understand the low energy charge transport properties of graphene one expands the full band structure close to the $\bs{K}$  point by writing  $\bs{q} = \bs{K}+\bs{k}$, with $ \bs{k} \ll \bs{K}$ and then by keeping the first order term and neglecting all higher order terms in $\bs{k}$, the resulting dispersion relation is of the form $E = \pm \hbar v_{F}|\bs{k}|$ such that low energy quasiparticles of graphene are described by the Dirac-like Hamiltonian \cite{szwski, semenoff, haldane}
\beq \bs{\hat{H}_{0}} = v_F(\bs{\sigma}\cdot \bs{p})  \label{graham} \eeq 
where $v_F\approx 10^{6}ms^{-1}$ is the fermi velocity, $\bs{\sigma} = (\sigma_x,\sigma_y)$ is the vector comprising of two component Pauli matrices, and $\bs
{p} = (p_x,p_y)$ is momentum vector in $x-y$ plane.  Thus, charge carriers in graphene behave like massless relativistic fermions dubbed as massless Dirac fermions with only the velocity of light is replaced by the Fermi velocity. This way, the behaviour of transport electrons in graphene is very different from those in ordinary semiconductors where they have a parabolic dispersion like a non-relativistic free particle. In the following sections we shall describe how such charge carriers
with ultra relativistic dispersion laws transports in presence of highly inhomogenous magnetic field dubbed as magnetic barriers in combination with local gate voltages. 

\section{Electron transport in graphene : Optical analogy} \label{scalarbarrieroptics}

\subsection{Electron optics in the presence of scalar potential barriers}

In a two dimensional electron gas (2DEG) it is well-established that transmission of de Broglie waves satisfying the Schr$\ddot{o}$dinger equation through a one-dimensional electrostatic potential is similar to light propagation through a refractive medium. Such transport can be understood in terms of phenomena like reflection, refraction and transmission, leading to an analogy between electron transport and light propagation \cite{SIVAN90, Stormer, dutta}.
When a non-relativistic electron in a 2DEG  with quadratic dispersion, at Fermi energy $E$,  is incident on a potential barrier $V$, its momentum parallel to the interface outside and inside the barrier is conserved; i.e., $|\vec{p}_1| \sin \theta_1 = |\vec{p}_2| \sin \theta_2$, where $\vec{p}_{1,2}$ are the momenta and  $\theta_{1,2}$ are the angles they make with surface normals respectively in the regions without and with potential barrier. This leads to the following Snell's law:
\beq \frac{\sin \theta_1}{\sin \theta_2} = \left(1-\frac{V}{E}\right)^{\frac{1}{2}} \label{electronwave1} \eeq

Graphene charge carriers however do not have the quadratic dispersion, but instead behave as massless Dirac-Weyl fermions leading to a different set of transport phenomena \cite{KSN1, KSN2, YZ1, Geimreview, Castroneto, Beenakker}. The optical analogues of such electron transport can again be constructed by considering the charge carriers in MLG incident on an electrostatic potential barrier $V$. Such charge carriers in monolayer graphene (MLG) obey 
the following Dirac-Weyl  like equation:
\beq v_F\left( \bs{\sigma} \cdot \bs{p}\right) \Psi(x,y) = \left(E-V_i \right)\Psi(x,y) \label{hamelecpot} \eeq
where $V_i = V_0$ in the barrier region ($ |x| \le d$) and vanishes outside, and $\Psi(x,y) = [\Psi_1, \Psi_2]^{T}$ is the two component wavefunction with $T$ denoting the transpose of the row vector.  We assume $V=V(x)$ to  be the one dimensional potential.  Due to the y invariance of potential barrier we consider the solutions of the form $\Psi(x,y) = \psi(x)e^{ik_y y }$. On substituting this in Eq.(\ref{hamelecpot}), two coupled equations in $\psi_1$ and $\psi_2$ are obtained as:
\beq -i\left[ \frac{d}{d x} \pm k_y \right]\psi_{2,1} = (E-V_i)\psi_{1,2} \nonumber \eeq

The above two off-diagonal equations can be decoupled in terms of $\psi_{1,2}$ to give a  Schr$\ddot{o}$dinger like equation of the form 

\beq \left[ -\frac{d^2}{d x^2} + k_y^2 \right]\psi_{1,2} = (E-V_i)^2\psi_{1,2} \label{decopeqn} \eeq
which admits exponential solutions,namely $ \psi_{1,2} \propto \exp( - ik_x x)$. 

By parameterizing the energy momentum relation  $k^2_{x_i}+k_y^2 = (E-V_i)^2$ in polar coordinates, we obtain
\bea k_y = E\sin\phi,  &  k_{x_1} = E\cos\phi, ~~~ |x|>d, \nonumber \\
k_y = (E-V)\sin\theta,  & k_{x_2} = (E-V)\cos\theta, ~~~ |x|<d. \nonumber \eea

\begin{figure}
\begin{center}
\includegraphics[width=100 mm]{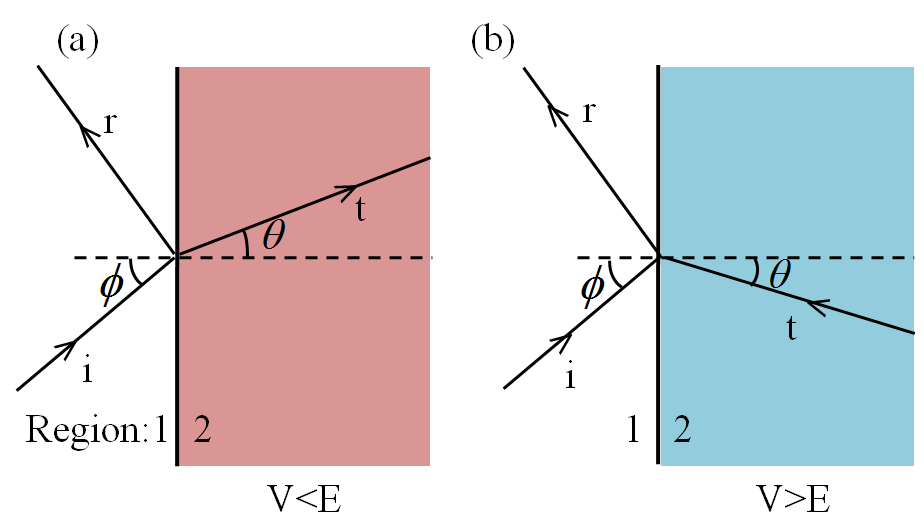}
\end{center}
\caption{Schematic of the transmission through a potential step for an electron with (a) $E>V$ (b) $E<V$.}
\label{splitgate}
\end{figure}

The momentum component along a straight interface should be conserved. Accordingly, using the electron momentum conservation in the y direction at the left interface $x = -d$, we obtain the Snell's law in the following form:
\beq (E-V)\sin\theta = E\sin\phi \nonumber \eeq
Clearly for $E<V$, the barrier acts like a medium with negative refractive index \cite{cheinov}. This is shown in Fig.\ref{splitgate} 
\beq n = \frac{\sin\theta}{\sin\phi} = -\frac{E}{|E-V|} \label{analogyeq1}\eeq
This corresponds to the  electronic analogue of the well known phenomenon of negative refraction, which occurs in left handed metamaterials  and which was first proposed by Veselago \cite{Veselago} and subsequently developed 
in more detail by Pendri and collaborators \cite{pend85}. If one uses so called split gate voltage by  placing the region $V < E$ and $V>E$
side by side ( Fig. \ref{splitgate}), the Fermi level can be tuned below and above the Dirac point. Since the region below and above the charge neutral Dirac points are respectively hole and electron  states for such Dirac fermions, this creates $p$ and $n$ type region in graphene simply
by changing the height of of the electrostatic potential barrier in a given region \cite{cheinov}. The resulting structure 
is called graphene $p-n$ junction. Consequently, a graphene $p-n$  junction can be used as a Veselago lense for electron focussing. 
Similarly circular graphene p-n junctions has also been studied showing the formation of caustics \cite{jozsef}.

\subsection{Klein Tunneling} \label{KTsection}

While electrostatic potential barriers can indeed manipulate charge transport in graphene, an electron could tunnel through any high barrier in contrast to the conventional tunnelling of non-relativistic electrons \cite{ando, KSN3}. This behaviour, called Klein tunnelling in graphene, leads to several observable transport effects related to transport some of which have been demonstrated in graphene \cite{kleinexp, kleinexp1} and also in carbon nanotubes \cite{Steele}. 
Below we review  briefly the work on such tunnelling problem for graphene charge carriers in the presence of an electrostatic potential barrier. This analysis will then be extended to similar transport through inhomogenous magnetic fields and will be followed by comparing and contrasting these transport to electrostatic and magnetic barriers, which is the central topic of this  review article.\

The wavefunction solutions in the presence of a potential barrier can be obtained by solving Eq.\ref{decopeqn}. These solutions in any region of space can be written in terms of linear combination of forward and backward propagating plane waves such that 

\beq \psi_1(x) =  \left \lbrace \begin{matrix}  e^{ik_x x} + r e^{-ik_x x},  &  x<-d  \\
 ae^{iq_x x} + b e^{-iq_x x},   &  |x|<d \\ 
  te^{ik_x x} ,  &  x>d \end{matrix} \right .\nonumber  \eeq

\beq \psi_2(x) =  \left \lbrace \begin{matrix} s\left(e^{ik_x x+i\phi} - r e^{-ik_x x-i\phi} \right),  &  x<-d  \\
s^{\prime}\left( ae^{iq_x x+i\theta} + b e^{-iq_x x-i\theta} \right),   &  |x|<d \\
ste^{ik_x x+i\phi} ,  &  x>d   \end{matrix} \right . \label{solutionelec}  \eeq

where $k_F = 2\pi/\lambda$ is the Fermi wavevector, $k_x = k_F \cos \phi$ and
$k_y = k_F \sin\phi$ are the wavevector components outside the barrier, $q_x = \sqrt{(E-V)^2/\hbar^2v_F^2 -k^2_y}$ ,$\theta = tan^{-1}(k_y/q_x )$ is the refraction angle, $s =sgn( E)$ and $s^{\prime} =sgn(E-V_0)$.

By taking into account the continuity of wave-function components $\psi_1$ and $\psi_2$ at the boundaries of the barrier the transmission and reflection coefficients are obtained as 
\bea t &=& \frac{2ss^{\prime}e^{-ik_xD}\cos\phi \cos\theta}{ss^{\prime}[e^{-iq_xD}\cos(\phi+\theta)+e^{iq_xD}\cos(\phi -\theta)]-2i\sin q_xD} \label{transformula} \\
r &=& \frac{2ie^{i\phi}\sin q_xD[\sin\phi-ss^{\prime}\sin\theta]}{ss^{\prime}[e^{-iq_xD}\cos(\phi+\theta)+e^{iq_xD}\cos(\phi -\theta)]-2i\sin q_xD} \label{rformula} \eea

\normalsize
Here $D = 2d$ is the width of the barrier. Clearly $q_xD = n\pi, ~~ n = 0, \pm 1, ... $ corresponds to the resonance condition at which the barrier transmission is unity. This corresponds to usual resonant tunnelling through a potential barrier and occurs for non 
relativistic electron as well.  However remarkably, the above expression also shows that barrier always remains transparent for normally incident carriers, ie. $T= 1$ when $\phi = 0$. This feature of anomalous tunnelling in graphene is peculiar to the Dirac like spectrum for graphene charge carriers and is explained as follows: As explained briefly in section \ref{general}, the gapless, conical spectrum of graphene is the result of intersection of two cosine like energy bands originating form sublattices A and B. (represented in green and red for A and B sublattice respectively, see fig.\ref{bandcone}). Due to this, the sublattice degree of freedom i.e. pseudospin should remain fixed on each branch. This also means that an electron with energy $E$ and possessing wavevector $\bs{k}$ will originate from the same branch as a hole with energy $-E$ and possessing wavevector $\bs{-k}$. Above two features allows the introduction of chirality for graphene charge carriers. The term $(\bs{\sigma}\cdot \bs{p})$ present in the Hamiltonian (\ref{graham}) gives the chirality, namely projection of pseudospin on the direction of motion, 
\beq \bs{C} = \frac{\bs{\sigma} \cdot \bs{p}}{|\bs{p}|} \nonumber \eeq and has two possible eigen values, $+1$ (for electrons)  and $-1$ (for holes).  This operator is same as the usual helicity operator for $3+1$-dimensional relativistic 
electrons that obey Dirac equation \cite{Greiner}, but redefined for charge carriers in graphene which obeys $2+1$ dimensional Dirac-Weyl equation.  
In the absence of any potential, the chirality operator commutes with the Hamiltonian and is therefore a conserved quantity. For normally incident carriers, the chirality remains conserved even in the presence of an external electrostatic potential $V(x)\mathbb{1}$. This can be shown as follows:
For normally incident carriers, 
\bea Chirality, \bs{C} &=& \frac{\bs{\sigma_x}p_x}{p_x} \nonumber \\
&=& \bs{\sigma_x} \nonumber \eea
Hence \beq [\bf{C},\hat{H}]  = [\sigma_x,k_x\sigma_x+V(x)\mathbb{1}] = 0 \nonumber \eeq
Also the velocity operator 
\beq \bs{v_x} = -i[x,\hat{H}] = \bs{\sigma_x} \nonumber \eeq

Thus the velocity operator is same as the chirality which is a conserved quantity for this case \cite{fuchs, todorovskiy}. The velocity along the $x$ direction is therefore a constant of motion and thus cannot be reversed. This leads to perfect transmission through such a barrier at normal incidence. This is what describes Klein tunnelling in MLG \cite{KSN3}. This is to be noted here that a rectangular electrostatic barrier in graphene assumes that the gate voltage-induced doping  changes abruptly at the edges . In reality, however, the doping level continuously changes and thus the edge of the potential  barrier is actually smooth and not sharp. This issue has been considered for scalar potential barrier in Ref.\cite{falko}, where it was found  that a potential which is smooth on the scale of the Fermi wave length for small angles of incidence,  $T(\phi)=\exp(-\pi k_{F} D \sin^{2} \phi)$ where $D$  is the barrier width. A comparison with the transmission expression given in Eq.\ref{transformula} shows that the assumption of the rectangular barrier captures the effect of Klein tunnelling correctly. However, at other  angles close to the normal incidence it overestimates the transmission. Transmission through trapezoidal barrier was also analysed in Ref.\cite{Sonin} which combines the effect of a smooth barrier and a rectangular barrier. \


On expanding $\phi$ in eq.(\ref{rformula}), close to the normal incidence, namely $\phi=0$, by substituting $\phi = \delta\phi_0$, it can be seen that the  reflection amplitude r undergoes a $\pi$ phase jump when the incident angle $\phi$ goes from positive to negative value . At zero magnetic field, at non-normal incidences, the two consecutive reflections on the two p-n interfaces occur with opposite angles $\theta_1 = \theta$ and $\theta_2 = -\theta $ (Fig.\ref{fabB}). 
When such a system is placed in a transverse magnetic field $\bs{B}=B \hat{z}$, the electronic trajectories bend in the presence of magnetic field. And above a critical field value $B > B_{c}$, trajectory bending becomes sufficient to make the two consecutive reflections occur with the same incident angle $\theta_1 = \theta_2$ which is exactly the case what happens at normal incidence in the absence of magnetic field. This suddenly adds $\pi$ to the phase accumulated by an electron between two reflections and shifts the interference fringes by half a period \cite{shytov}. The observation of this half-period shift in Fabry Perot interference fringes is therefore a direct evidence of perfect tunnelling at normal incidence \cite{kleinexp}. We analyse this in more detail in section \ref{FP}. An alternative way of demonstration of Klein tunnelling in graphene p-n junctions has been addressed in \cite{kleinexp1}. This was done by probing the transition from clean to disordered transport across a single steep p-n junction. Very recently, an angle dependent carrier transmission probability in graphene p-n junctions has also been investigated experimentally \cite{sutar} and theoretically \cite{avik} where it is shown that chiral tunnelling can be directly observed from the junction resistance of a tilted interface probed with separate split gates.


\section{Electron transport in the presence of inhomogenous magnetic field profile}
While the above mentioned absence of backscattering by a scalar potential due to Klein tunnelling is a very interesting phenomenon, this implies that confining transport electrons in graphene by a potential barrier is not possible in a conventional way. This leads to problem in device making. For example, even though it is easy to make a $p-n$ junction 
in graphene, reverse biasing  such  junction will be very difficult for Klein tunnelling.
 For the designing of graphene based electronics it is crucial to attain confinement of electrons within a mesoscopic or nanoscopic size of the sample. For this reason, several alternatives have been suggested. One way is to exploit the fact that suitable transverse states in a graphene strip may allow one to circumvent Klein tunnelling \cite{silvestrov}. Other schemes that have been proposed include gated nanoribbons \cite{trauzette}, gated \cite{milton} or doped  \cite{pereira} bilayer etc. Yet another possibility was demonstrated theoretically in \cite{eggerprl},\cite{eggerssc} by making use of external inhomogeneous magnetic fields  applied perpendicular to the graphene plane. This brings us to the the main focus of this review article- to investigate the electron transport in the presence of inhomogenous magnetic field viz. magnetic barrier(s) in graphene. Before proceeding further we first discuss the experimental strategies for creating such inhomogenous magnetic fields and point out that such magnetic barriers are already in very much use in other materials. 

\subsection{Producing inhomogenous magnetic field profile: Experimental strategies}\label{magexpt}
A  microscopically inhomogenous magnetic field can be created for charge carriers in graphene by placing a graphene sheet in close proximity to long magnetic stripes that produce highly localized magnetic fields. Such field profiles can be generated using demagnetizing fields produced at the edges of narrow stripes made with hard ferromagnetic (FM) materials of either perpendicular or in-plane anisotropy. The magnetic field of such structures is given as 
\beq \bs{B}=B(x,z_0)\hat{z}=B_0[K(x+d,z_0)-K(x-d,z_0)]\hat{z} \nonumber \eeq
where $K(x,z_0)=-\frac{4xd}{x^2 + z_0^2}$ for perpendicular magnetization, and $K(x,z_0)=-\frac{2z_0d}{x^2 + z_0^2}$  for magnetization parallel to the graphene sheet. $B_0$ is a constant dependent on the aspect ratio of the stripe.  
\begin{figure}[t]
 \begin{center}
\includegraphics[width=100 mm]{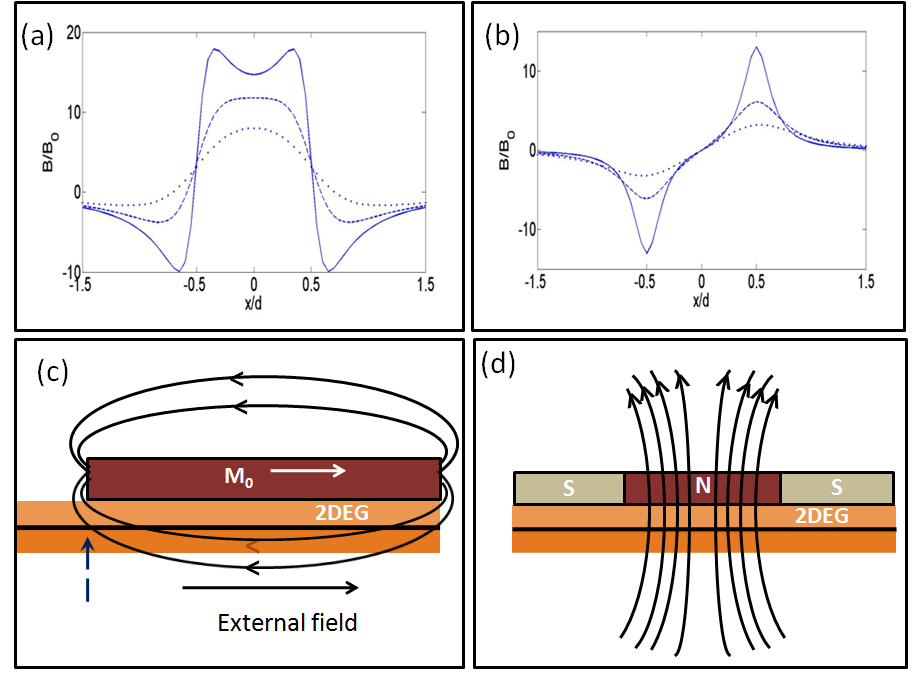}
 \end{center}
 \caption{Magnetic field under the stripe corresponding to the (a) perpendicular (b) parallel configurations.The magnetic field is given at the following distances from the magnetic stripe: $z_0 = 0.15$ (solid curve), $z_0 = 0.3$ (dashed curve), and $z_0 = 0.5$ (dotted curve).}
\label{peetersfiglabel}
\end{figure}

For a given value of $z_0$ we plot in Fig.\ref{peetersfiglabel}(a), (b) the profile of such a magnetic field. The major component of the demagnetizing field reaches the graphene monolayer and produces the desired field profile. Though there is always some component of the demagnetizing field that will give rise to undesired fringe fields in other directions, these can be substantially lowered by suitable magnetic design of the stripes. Such nanostructures are routinely used in magnetic recording media \cite{appphys}. Materials such as CoCrPt  produce fields of 1 Tesla close to the surface with bit lengths ranging from 50-100 nm.  For example, in \cite{IEEE}, isolated tracks of single-domain magnetic islands have been fabricated using focused ion-beam lithography for both perpendicular and parallel anisotropy using CoCrPt.  Here, patterns with successive magnetizations pointed along opposite directions were achieved. In another recent work, off-axis electron holography has been used to probe the magnetization structure in high density recording medium by using perpendicular magnetic anisotropic (PMA) recording medium \cite{dipole}. The direct imaging of magnetization done shows that the foils of PMA material consist of successively reversed highly stable domain structures of few ten's of nanometer size. In practice, one can also change the strength of the magnetic field by suitable adjusting the width of such PMA material. Precise design of read-write structures for recording individual bits at these dimensions has also been achieved \cite{nphotonics}. Using the above discussed techniques, typical magnetic barriers can be patterned down to 50-100 nm widths. \

\begin{figure}[t]
    \begin{center}
\includegraphics[width=100 mm]{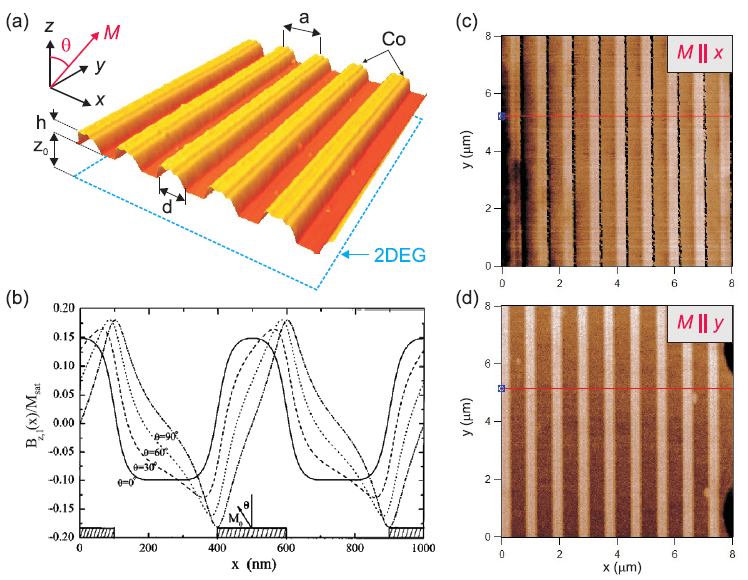}
    \end{center}
    \caption{{\it (color online)}(a) Cobalt finger gates at the surface of a 2DEG. (b) Magnetic modulation profiles at different tilt angles of the magnetization.(c) MFM image of the cobalt grating magnetized along the short axis of the stripes, in the plane of the 2DEG. The magnetic poles appear as the black lines. (d) MFM image of the grating magnetized along the long axis. The magnetic poles are absent. Parameters: $a = 400 nm$, $d = 200 nm$, $h = 160 nm$ and $z_0 = 90 nm$. [This figure is taken from \cite{nogaretjpcm}, printed with permission.]}
\label{nogaretfig}
\end{figure}

Apart from the above mentioned method of using ferromagnetic materials \cite{mancoff} there are several other ways in which inhomogenous magnetic fields have been experimentally realised in conventional semiconductors. One such way is through the integration of superconducting elements,  which may be used to screen the externally applied magnetic field  in accordance with Meissner effect. When an external magnetic field is applied the flux lines will be expelled from the superconductor due to the Meissner effect. If the latter is close enough this will result in an inhomogeneous magnetic field in the 2DEG \cite{vonklitzing1}\cite{geimjetp} . This is shown schematically in [Fig.\ref{peetersfiglabel}(d)]. As an example, in ref. \cite{vonklitzing1}, it was shown that  with type II superconducting films deposited on top of the two-dimensional electron gas in a GaAs/AlGaAs heterostructure, the distribution at the 2DEG takes the form of flux tubes which are much narrower than an electron inelastic scattering length. The superconducting materials used in these experiments were typically 200nm of lead (Pb), 400 nm of a lead/indium [Pb(1 at. $\%$ In)] alloy, and 200 nm of niobium nitride (NbN), as being materials whose characteristic superconducting length scales span a broad range. Growing the heterojunction on top of a pre-etched (nonplanar) substrate can also give rise to inhomogeneous magnetic fields \cite{leadbeater}-\cite{grayson}. When a uniform external field is applied on such a system, the angle between the field direction and the normal to the 2DEG depends on the tilt of the facet and therefore will have a normal component of magnetic field which varies spatially across the sample. Since it is only the normal component of magnetic field that influences the transport, by varying the facet length and angle, as well as the angle between the applied field and the normal to the substrate, a wide variety of field profiles can be generated.\

In other studies\cite{apl6}-\cite{kubrak} a similar yet simpler approach was used based on the idea that if a thin magnetic film is placed on top of a heterostructure its in-plane magnetization can be saturated in an external in-plane magnetic field [Fig.\ref{peetersfiglabel}(c)]. In this case the out-of-plane component of the fringe field under the edge of the film creates a magnetic barrier for electron transport with a width of the order of 100 nm determined by the separation of the 2DEG from the magnetic film. Magnetic field strengths of more than 0.5 T have been realized in this way \cite{kubrak}.
A detailed discussion on experimental methods used for applying magnetic modulations to 2DEGs can be found in the review article \cite{nogaretjpcm}.


\subsection{Magnetic confinement of massless dirac fermions in graphene} \label{finitemagbarsect}
In the previous section we reviewed the experimental progress of realizing inhomogenous magnetic field or magnetic 
barrier over a wide range of length scales in a number of systems. 
Mesoscopic transport in presence of  inhomogenous magnetic fields has been studied 
theoretically \cite{soumaphyreports}, \cite{peetersprb48} - \cite{ibrahimpeetersprb52} for non relativistic 
electrons .  Related experimental studies also took place in conventional semiconductor heterostructures, e.g., transport in the presence of magnetic barriers \cite{johnson} and superlattices \cite{carmona}, magnetic edge states close to a magnetic step \cite{nogaret}, and magnetically confined quantum dots or antidots \cite{dubonos}. In the following section B we review the transport of massless Dirac fermions in graphene in the presence of of magnetic barriers with some suitable 
examples.
 
The magnetic barrier profile chosen here is in the form of step like function  \cite{eggerprl,eggerssc} as 
\beq  \textbf{B} = B\bs{\Theta}(d^2-x^2)\hat{z} \eeq, 
 and the vector potential chosen in Landau gauge $A_y(x) = [-Bd, Bx, Bd]$  in region $x<-d$, $|x|<d$ and $x>d$ respectively.

\begin{figure}
\begin{center}
\includegraphics[width=100 mm]{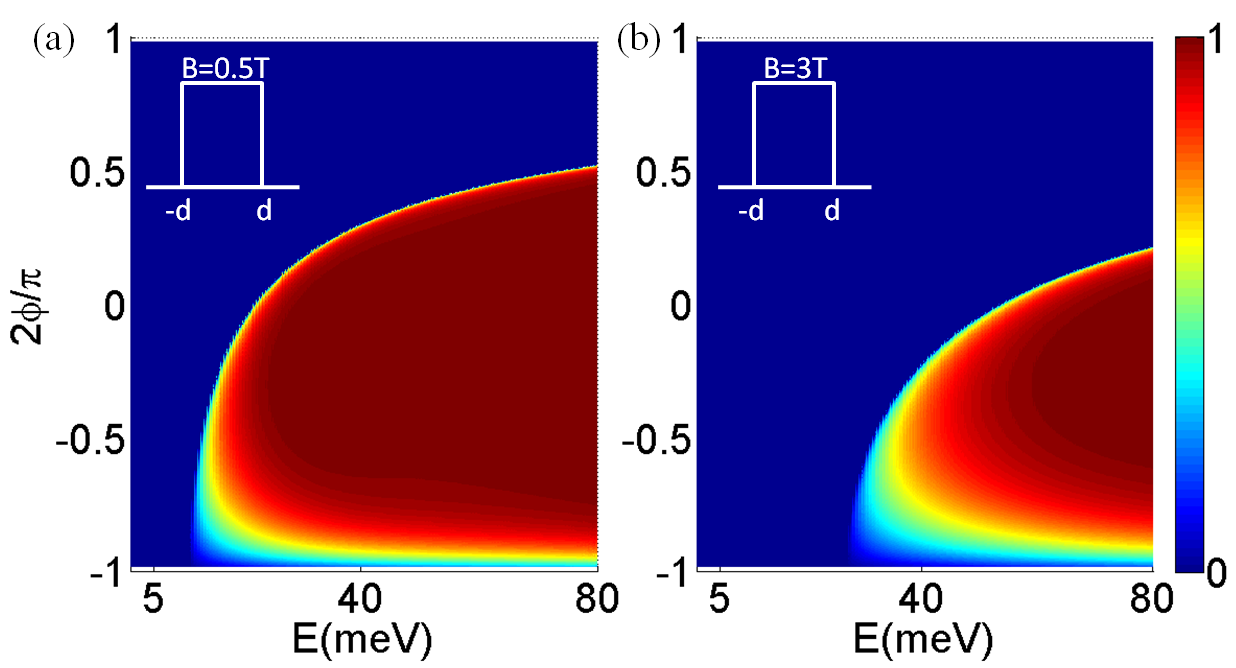}
\end{center}
\caption{Transmission in the presence of finite width magnetic barrier in monolayer graphene, $D = 2d = 1.5l_B $ (a)$B = 0.5T$ (b) $B = 3T$.}
\label{finitemagbarlab}
\end{figure}

Typically, the equation of motion of charge carriers at the  two inequivalent Dirac points K and $K^{\prime}$  are decoupled from each other \cite{shon} in the absence of intervalley scattering which gives a wave vector change of the order of $2k_{F}$. We assume for the cases to be considered this condition is satisfied.
Under this situation motion near each Dirac point  can be treated separately. The stationary solutions  near each such point is given by a two component spinor and satisfies the equation
\beq v_F(\bs{\pi_x}\mp i\bs{\pi_y})\Psi_{2,1} = E\Psi_{1,2} \label{ham1}\eeq
Here, $v_F$ is the fermi velocity ($v_F\approx c/300$) and $\bf{\pi} = \bf{p}+\frac{|e|}{c}\bs{A}$. Using $l_B$ as the unit of length scale such that $\bar{x} = \frac{x}{l_B}$and $\frac{\hbar v_F}{l_B}$ as the unit energy such that $\epsilon = \frac{El_B}{\hbar v_F}$ and $\Psi(x,y) = \psi(x)e^{ik_y y}$ in the Landau gauge, Eq.\ref{ham1} can be rewritten as
%

\beq -i \left[\frac{\partial}{\partial \bar{x}} \pm \left(k_yl_B+ \frac{A_y(x)}{Bl_B}\right) \right]\psi_{2,1} = \epsilon\psi_{1,2} \label{coupledfinite}\eeq
which can be decoupled to obtain the following Schr$\ddot{o}$dinger like equation:

\bea  \left [-\frac{\partial^2}{\partial\bar{x}^2}+\left (k_yl_B+\bar{x}\right )^2+1 \right ]\psi_{1,2}= \epsilon^2\psi_{1,2}  & |x|<d \nonumber \\
 \left [-\frac{\partial^2}{\partial\bar{x}^2}+\left (k_yl_B \mp d/l_B\right )^2\right ]\psi_{1,2}= \epsilon^2\psi_{1,2} \eea 
where $(-/+)$ sign in the above expression corresponds to region $(x<-d)/(x>d)$ respectively.
 
Upon solving, we obtain the solutions outside the barrier as propagating solutions while inside the barrier the solutions are localised and can be written in terms of parabolic cylindrical functions \cite{gradshteyn}. 

\beq \psi_1(x) = \left \lbrace \begin{matrix} e^{ik_x x} + r e^{-ik_x x},  &  x<-d  \\
 a D_{\epsilon^2/2-1}(\sqrt{2}(\bar{x}+k_yl_B)) &  \\
+ b D_{\epsilon^2/2-1}(-\sqrt{2}(\bar{x}+k_yl_B)),  &  |x|<d  \\
te^{ik_x^{\prime} x} ,  &  x>d  \end{matrix}  \right . \label{psi1loc} \eeq

\beq \psi_2(x) =   \left \lbrace  \begin{matrix} e^{i(k_x x+\phi)} + r e^{-i(k_x x+\phi)},  &  x<-d   \\
 a\frac{i\sqrt{2}}{\epsilon} D_{\epsilon^2/2}(\sqrt{2}(\bar{x}+k_yl_B)) &  \\
+ b \frac{-i\sqrt{2}}{\epsilon}D_{\epsilon^2/2}(-\sqrt{2}(\bar{x}+k_yl_B)),  &  |x|<d \\
 te^{i(k_x^{\prime} x+\phi^{\prime})}   &  x>d  \end{matrix}  \right .  \label{psi2loc} \eeq

where  $k_x =k_F\cos\phi $,  $k_x^{\prime} = k_F\cos\phi^{\prime}$,  $D$ stands for parabolic cylindrical function, and $\phi$ and $\phi^{\prime}$ corresponds to the incidence and emergence angle in regioin I and III respectively.

The emergence angle can be obtained by conservation condition of $k_y$ as:
\beq \sin\phi^{\prime} = \frac{2d}{\epsilon l_B}+ \sin\phi \eeq
An alternative way of writing the above equation 
is \beq \epsilon \sin \phi' = 2\pi\frac{\Phi}{\Phi_{0}} + \epsilon \sin \phi \nonumber \eeq 
where  $\Phi_0 = hc/e$ and $\Phi = 2Bd l_{B}$, is the net flux enclosed by an area $2d l_{B}$.
The second form particularly  shows how the flux enclosed by the area containing the barrier changes the 
refractive index of this region in a non specular way.  

The transmission through such magnetic barrier can be obtained by matching the wavefunction  at the boundaries of the barrier, $T= \frac{k_x}{k_x^{\prime}}|t|^2 $, where the prefactor ensures the current conservation in the system. The transmission as function of incidence angle $\phi$ and incident energy $E$ is shown in fig.\ref{finitemagbarlab}. As one can see, Eq. \ref{psi1loc} and \ref{psi2loc} depicts the formation of localised states inside the barrier region. However these states are bound only in the direction perpendicular to the barrier but can propagate in the direction along the inhomogeneity of the magnetic field. The transmission, for this reason, depends very strongly on the incident wavevector and facilitates the possibility to construct wave-vector filter \cite{Masir1}. Fig. \ref{finitemagbarlab} depicts this situation clearly. The requirement of the propagating solutions in the incident region and region of exit determines the range $ -\pi/2<\phi<\sin^{-1}\left(1-\frac{d}{\epsilon l_B}\right) $ beyond which no transmission is possible, thereby leading to the possibility of confinement. This is how magnetic barrier turn reflective for a set of wave vector for the massless Dirac fermions. In the following section we shall illustrate this property more clearly by building clear optical analogy of 
electron transmission through the extremely inhomogeneous magnetic barrier where the magnetic field profile 
can be approximated as a delta function. 

\section{Massless dirac fermions in inhomogenous magnetic fields: An analogy with light propagation}
\subsection{Magnetic vector potential (MVP) barrier}\label{MVPsection}
The issue addressed here is whether the transport through such magnetic barrier can be understood in terms of 
propagation of light through medium with changing dilelectric constant using well known ideas in geometrical optics\cite{sgms}. Even though such idea was developed for the ballistic electron transport through electrostatic 
potential barrier, generalizing this concept in presence of a magnetic field was non trivial since in the presence of magnetic 
field electron executes cylotron motion and the wave vector $\bs{k}$ is not a good quantum number for such motion.
Thus to yield a Snell's law like one in Eq. (\ref{analogyeq1}) in  presence of a magnetic barrier, it is essential that the electron should not complete its half cyclotron radius beyond which it get completely localised inside the barrier. This situation can be realized in the presence of a highly inhomogeneous magnetic field. Particularly, if the range of inhomogeneity is much smaller than the cyclotron radius, one is left with plane wave-like scattering states.  Thus, such field profile scatter the electrons in the same way as an electrostatic potential does.

For this to be valid, two conditions must be satisfied. First, the magnetic length $l_B = \sqrt{\frac{\hbar c}{eB}}$ should be similar in order to the width of such magnetic barriers. Second, the de Broglie wavelength $\lambda_F$ should be much larger than the barrier width so that the electron will not see the variation in the vector potential inside the barrier. An extreme case of such an inhomogeneous magnetic field is singular magnetic barrier - the one introduced in \cite{peetersprl, ibrahimpeetersprb52, magb2, sgms, mssg, neetuinsa, Masir1}, which gives rise to step-function like magnetic vector potential (MVP) barriers. Here, we shall consider the scattering of massless Dirac fermions by such singular magnetic barriers as in Eq.\ref{mvpbarriereq}. Such field profiles having highly localized field variations are well known in literature and have been realized experimentally as we have already pointed out in the section \ref{magexpt}. Patterned stripes down to length scales as low as 10 nm have been realized using nanolithography \cite{terriskrawczykbader}. It is possible to achieve field profiles at even smaller dimensions using domain walls with widths in the range of 10-50nm and magnetic nanostructures down to 5 nm \cite{sunsci, zhengFePt} and 0.15 nm \cite{bergmann} having highly localized field variations which can 
be approximated as such delta function barriers. 

\subsubsection{Electron optics with MVP barrier}
\begin{figure}
\begin{center}
\includegraphics[width=100 mm]{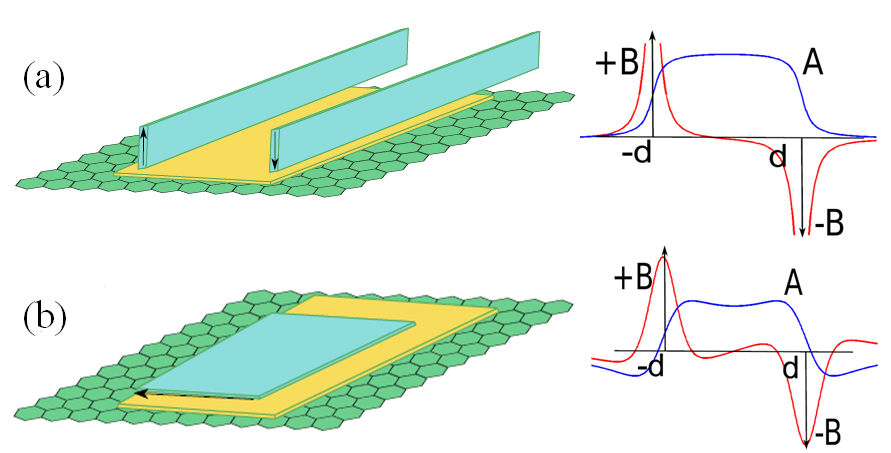}
\end{center}
\caption{Monolayer graphene with ferromagnetic stripes having magnetizations perpendicular (a) and parallel (b) to plane. The magnetic field B (red) produces a magnetic vector potential A (blue). Single MVP barriers are formed in (a) and (b).}
\label{profile}
\end{figure}

The most elementary of such  magnetic field profile that satisfy the flux line closure condition and its vector potential in the Landau gauge can be given as :
\bea \textbf{B} &=& Bl_B[\delta(x+d)-\delta(x-d)]\hat{z} \label{mvpbarriereq} \\
\bs{A}_y(x) &=& Bl_B\Theta(d^2-x^2)\hat{y} \nonumber \eea
In the absence of intervalley scattering, the charge carriers in graphene in the presence of above potential form  can be described by Eq.\ref{ham1} which can be rewritten in the form of the following coupled equation
\beq -i \left[\frac{\partial}{\partial \bar{x}} \pm(k_yl_B+\Delta) \right]\psi_{2,1} = \epsilon\psi_{1,2} \label{coupled}\eeq
Here $\Delta = 1$ for $\left|x\right|<d$ and $\Delta = 0$ for $\left|x\right|>d$. The coupled equations Eq.\ref{coupled} can be decoupled easily and results in a Schr$\ddot{o}$dinger like equation of the form 
\beq \left[-\frac{\partial^2}{\partial \bar{x}^2} + (k_yl_B+\Delta)^2 \right]\psi_{1,2} = \epsilon^2\psi_{1,2} \label{uncoupled} \eeq
Thus, in the barrier region $-d<x<d$ electrons see a momentum-dependent barrier of height $[k_y-sgn(e)\frac{1}{l_B}]^2$. Hence such a barrier will be termed as  magnetic vector potential barrier (MVP). 

\begin{figure}[t]
\begin{center}
\includegraphics[width=100 mm]{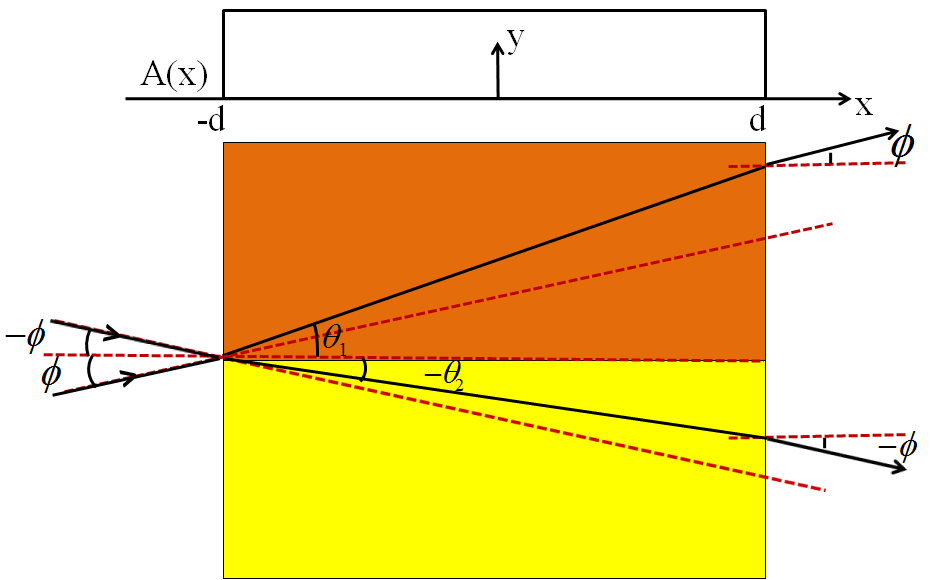}
\end{center}
\caption{Asymmetric refraction through a single MVP barrier.}
\label{analogyplot}
\end{figure}

Since magnetic field doesn't do any work, the energy conservation gives $k_x^2 + k_y^2 = k_F^2$ for $\left|x\right|>d$ and $q_x^2 + (k_y+\frac{1}{l_B})^2 = k_F^2$ for $\left|x\right|<d$.
By parameterizing the above two equations in  polar co-ordinates as shown in Fig. \ref{analogyplot}, we obtain the relation 
\beq sin\left|\theta\right| = sin\left|\phi\right| + sgn(\phi)\frac{1}{k_Fl_B}, -\pi/2<\phi<\pi/2 \label{analogy}\eeq
The situation is depicted in Fig.\ref{analogyplot}.  For a wave incident with positive $ \phi $ wave vector bends away from the normal whereas for a wave incident with negative incidence angle the corresponding wave vector bends towards the surface normal inside the barrier region.  Here, $\theta$ is the angle of refraction (denoted as $\theta_{1,2}$ for positive and negative angles of incidence). Clearly, the Snell's law for electron waves in such magnetic barriers is not specular as it is for light waves on smooth surfaces or for the incidence of the electrons on an electrostatic potential barrier \cite{cheinov}, \cite{KSN3}.

According to Eq.\ref{analogy}, when $\left|sin\left|\theta\right|\right| > 1$, $\theta$ becomes imaginary and the wave in the second medium becomes evanescent. In the language of optics this corresponds to total internal reflection (TIR).  According  to Fig. \ref{mvp}, this will happen when $\sin |\theta| > 1$
for  $0 < \phi \le  \frac{\pi}{2}$  and  when $\sin |\theta| < -1$  for  ${-\frac{\pi}{2}} \le \phi < 0$. In the latter case, this requires the wave vector to be negatively refracted \cite{falko} at sufficiently high magnetic field before TIR occurs. It also follows that for a given strength $B$  the magnitude of critical angle of  incidence $|\phi|=\phi_c$ for TIR is  higher for $ -\frac{\pi}{2} \le \phi < 0 $ as compared to the one for  $0 < \phi < \frac{\pi}{2}$. Because of TIR the transmission on both sides of Fig \ref{mvp} drops to $0$ beyond a certain value of $\phi$ and for fixed B field this value is higher for negative angles of incidence . Thus we can understand  the reflective nature of such barrier using the corresponding Snell's law given 
in \ref{analogy}. To illustrate this we shall now calculate the transmission through such barrier. 

\subsubsection{MVP barrier: Transmission probability}
\begin{figure}[t]
\begin{center}
\includegraphics[width=100 mm]{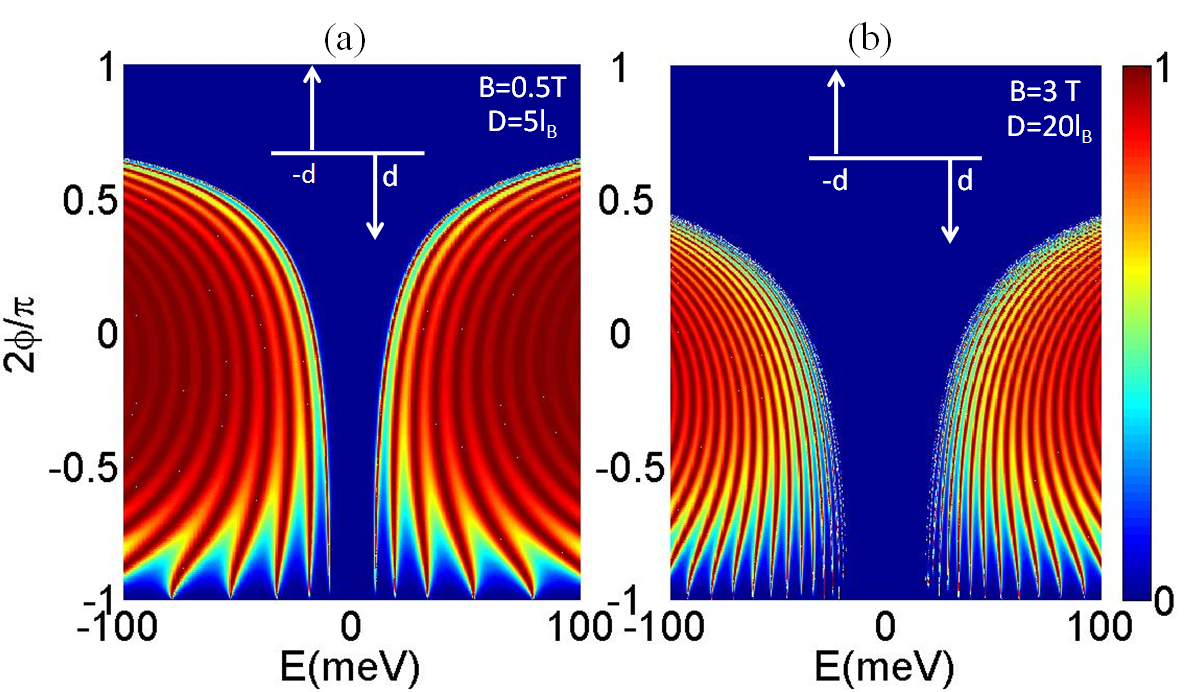}
\end{center}
\caption{Transmission T as a function of $\phi$ and $E$ for single MVP barrier (a) $B = 0.5T$ , $d=5l_B$ (b) $B = 3T$ , $d=20l_B$.}
\label{mvp}
\end{figure}

On solving Eq.\ref{uncoupled}, the corresponding wavefunctions in any region of space can be written in terms of a linear superposition of forward and backward moving plane waves. Then continuity of the wavefunction at the boundaries of the MVP barrier can be used to calculate the transmission coefficient as 
\small
\beq t = \frac{2ss^{\prime}e^{-ik_xD}\cos\phi \cos\theta}{ss^{\prime}[e^{-iq_xD}\cos(\phi+\theta)+e^{iq_xD}\cos(\phi  - \theta)]-2i\sin q_xD} \label{trans} \eeq
\normalsize
Here $D = 2d$. And s and $s^{\prime}$ are given by sgn($\epsilon$) and are both +1 for electrons when only magnetic fields are present. It may be pointed out that the form of the transmission expression is same as the corresponding one for 
electrostatic potential barrier (\ref{transformula}). But because the wave vector inside the barrier is different as compared 
to those in presence of scalar barrier, the transmission profile is fundamentally different. 
 Fig.\ref{mvp} shows how the Transmittance, $T = t^{*}t$ changes with the angle of incidence. As suggested by Eq.\ref{analogy}, the transmission is clearly asymmetric unlike the case of electrostatic potential \cite{KSN3}. \

For high electrostatic barriers such that $V>>E_F$, the wavevector is given by $q_x = \sqrt{\frac{(E_F-V)^2}{\hbar^2v_F^2}-k_y^2}$, which is real. The corresponding transmission for electrostatic potentials is 
\beq T = \frac{\cos^2\phi}{1-\cos(q_xD)\sin^2\phi} \nonumber \eeq
and is 1 at $\phi = 0$, exhibiting Klein tunnelling for massless Dirac fermions \cite{KSN3}. In contrast at high values of the magnetic barrier since $\frac{1}{k_Fl_B}\propto\sqrt{B}$, the wavevector given by $q_x^2 = k_F^2-(k_y+\frac{1}{l_B})^2 = -\kappa^2<0$. This leads to TIR and not Klein tunnelling. As seen in Fig. \ref{mvp}, the magnitude of critical angle beyond which TIR occurs is lower for a higher magnetic field. Thus, a stronger MVP barrier leads to higher reflections as opposed to complete transmission at normal incidence by high electrostatic potential barrier.
Complete transmission occurs only for $q_xD = n\pi$ as given by Eq.\ref{trans}. This corresponds to resonant tunnelling for Dirac electrons and happens in the same way as for nonrelativistic electrons, appearing as a number of peaks in the plots in Fig.\ref{mvp} . The number of tunnelling peaks increases with barrier width for both MVP and electrostatic barriers.\

In conclusion, we see that the reflectance can be controlled by suitably modifying the strength and locations of the magnetic barriers and thereby changing the refractive index of the intervening medium in a novel manner. This principle could be the basis of structures like magnetic waveguide, where the reflection must be high at desired propagation angles, here, this could be manipulated by changing the magnetic field. Also for strutures like the resonant cavity, high reflection is needed near normal incidence. Geometries such as three-mirror or four-mirror cavities \cite{sgms} could be used for high reflection at other angles .

\subsection{Combined MVP barrier and electrostatic potential (EMVP) barrier} \label{EMVP}

The charge carriers in graphene have a linear band structure albeit only close to the Dirac point, coincident with the Fermi level $E_F$ in undoped graphene. Small electrostatic potentials greatly affect electron states by shifting the Dirac point with respect to $E_F$ and causing the graphene sheet to behave as either an electron-deficit (p-type) or a hole-deficit (n-type) material. This situation was also experimentally verified by observing electron hole puddles \cite{yacobi, Youngreview}
near the Dirac point in monolayer graphene.
Thus, the effect of electrostatic potentials on any proposed graphene structure must be included. \

Quantum hall effect in gate controlled p-n junction in graphene in presence of uniform magnetic field was already studied experimentally \cite{marcusscience} which reveals new quantum hall plateaus. The top gate geometry was utilized in controlling the edge channels in the quantum Hall regime and with control over local and global carrier density. The effect of local modulation of charge density and carrier type in graphene field effect transistors using a double top gate geometry has also been studied \cite{mandar}. Also, in a recent experiment by S. Pisana et al.\cite{Pisana}, the enhanced magneto-resistance of a monolayer graphene sheet has been measured by connecting it to two voltage and two current terminals and simultaneously exposing it to various magnetic field strength at room temperature. The differential voltage as a function of the magnetic field has been plotted thereby analysing the joint effect of  magnetic field and the applied voltage on the magneto-transport particularly close to the Dirac point. From the analysis of the magneto resistance data it was inferred that the band structure of transport gets strongly modified in presence of voltage and magnetic field. The above experiment result clearly shows that simultaneous application of voltage and magnetic field strongly influences band structure, as we'll also see in the following section.  For most of the above experiments the magnetic field applied is homogenous over the typical size of the sample. A complimentary case  where as for our case we consider magnetic barrier which is a highly inhomogenous magnetic field over the typical size of the sample.

\begin{figure}[ht!]
\begin{center}
\includegraphics[width=100 mm]{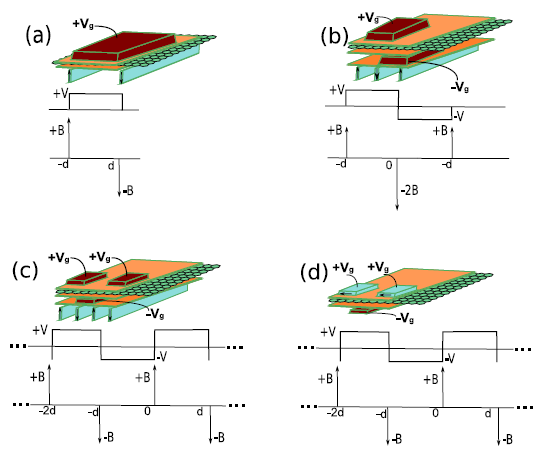}
\end{center}
\caption{ EMVP barrier structures for graphene. In (a)-(c), the magnetic field is applied using patterned ferromagnetic (FM) lines with perpendicular anisotropy and two potentials $+V_g$ and $-V_g$ are applied by separate conductor lines. As given in the text, effective potential induced in the graphene is $V$. In (b), the magnetic strength of the middle line is doubled by either using a different FM material or by larger dimensions. In (d), the magnetic field is produced from the two edges of a FM stripe with in-plane anisotropy and the same stripe is also used to apply  one of the potentials.}
\label{ferromag}
\end{figure}

We first discuss briefly some of the practical issues when a electrostatic gate potential is applied to the graphene sheet. A gate voltage $\pm V_{g}$ can be applied using a metal electrode and separating the electrode from the graphene layer with an insulating oxide layer. This oxide layer which serves as a dielectric medium between graphene sheet and metal electrode can be made as thin as several nanometers.  When such a voltage $V_{g}$ is applied locally, it induces electron (hole) doping $\pm \sigma_n$ proportionally and changes the carrier concentration within the channel. This shifts the undoped Fermi level $E_F$ from the Dirac point by an amount $V = sgn(\sigma_n) \sqrt{\sigma_n}\hbar v_F$, where $sgn(\sigma_n)$ is the sign of the induced charge. Due to this a local potential barrier V is created which is equal to the difference between the local Fermi level and the Fermi level in the undoped region.
Since the typical breakdown strength of dielectrics such as alumina and fused silica are around $10-20 MV/m$, a graphene layer could be subjected to 1V applied across a 100nm thick dielectric. Following Refs.\cite{KSN1,KSN2}, gate voltages $V_g$ of upto $\pm100 V$ have already been applied to graphene flakes.  The electrostatic gate potentials would be generally applied by separate conductors placed suitably in different planes than the FM stripes, but in some cases, the FM stripes could also be used to apply voltages.

\subsubsection{Electron optics with EMVP barrier}

\begin{figure}[ht]
\begin{center}
\includegraphics[width=100 mm]{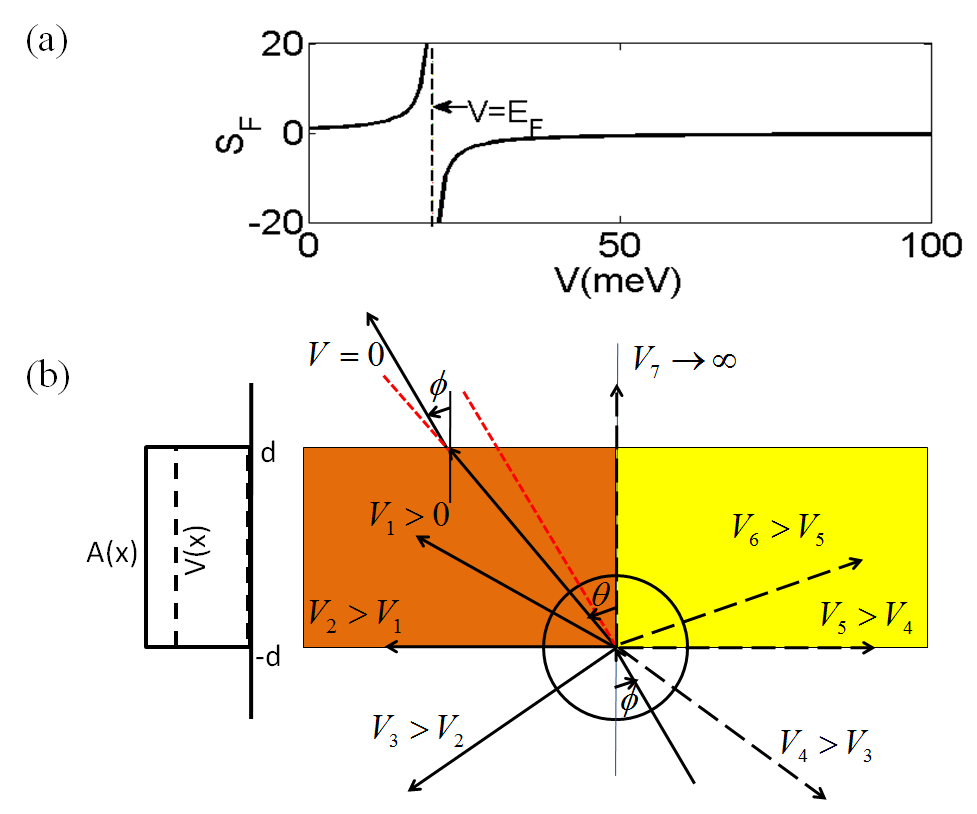}
\end{center}
\caption{(a) Function $S_F$ as a function of voltage (b) Ray diagram of electron propagation through a single EMVP barrier. With increasing $V$, the refracted angle $\theta_2$ increases continuously in the clockwise direction. Dashed rays are negatively refracted waves.}
\label{raydiag}
\end{figure}

Below we discuss the transport in the presence of a single MVP barrier and a commensurate electrostatic step potential (EMVP) barrier.

\bea V(x) = V, A_y(x) = Bl_B, \left|x\right|<d \nonumber \\ V(x) = 0, A_y(x) = 0, \left|x\right|>d  \eea
We assume $V>0$ and set the incident energy at $E = E_F$, where $E_F$ is the Fermi level located at the charge-neutral Dirac point in monolayer graphene and the voltage V is measured with respect to $E_F$. The entire treatment that follows is applicable in the neighbourhood of the charge-neutral Dirac points as long as the dispersion remains linear. In addition, we also assume that scattering by an EMVP barrier is not strong enough to break the degeneracy of the K and $K^{\prime}$ points. In the region $-d < x < d$ at either of the K points, the motion will be described by
\beq v_F \left(\begin{matrix} V/v_F & \bs{\pi}_x-i\bs{\pi}_y \\ \bs{\pi}_x+i\bs{\pi}_y & V/v_F \end{matrix}\right)\left(\begin{matrix}\psi_1 \\ \psi_2 \end{matrix}\right) = E\left(\begin{matrix}\psi_1 \\ \psi_2 \end{matrix}\right) \label{ham}\eeq

Here $\bs{\pi}=\bs{p}+\frac{e}{c}\bs{A}$. In the Landau gauge, the stationary solutions can again 
be written as 
\beq \psi_{1,2}(x,y) = \psi_{1,2}(x)e^{ik_y y} \nonumber \eeq
Substituting this in Eq.\ref{ham}, we get the following coupled equations
\small
\beq \begin{bmatrix} 0 & -i\partial_x-i(k_y+\frac{1}{l_B}) \\ -i\partial_x+i(k_y+\frac{1}{l_B})& 0 \end{bmatrix}\begin{bmatrix}\psi_1 \\ \psi_2 \end{bmatrix} = \frac{E-V}{\hbar v_F}\begin{bmatrix}\psi_1 \\ \psi_2 \end{bmatrix} \eeq
\normalsize
The above equation can be decoupled to yield 
\beq \left[-\partial_x^2+(k_y+\frac{1}{l_B})^2 \right]\psi_{1,2} = \left(\frac{E-V}{\hbar v_F}\right)^2\psi_{1,2} \label{decopemvp} \eeq 

The corresponding stationary solutions $\psi_{1,2}(x)$  are 
\beq \psi_1 = \left\{ \begin{array}{rl} e^{ik_x x} + r e^{-ik_x x} & \text{if }~~ x < -d\\ a e^{iq_x x} + be^{-i q_x x}  & \text{if }~~ |x|<d \\ t e^{i k_x x} & \text{if}~~ x > d \end{array} \right. \label{particle1}\eeq 

\beq \psi_2 = \left\{ \begin{array}{rl} s[e^{i (k_x x +  \phi)} - r e^{-i(k_x x + \phi)}] & \text{if }~~ x < -d\\ s'[a e^{i(q_x x + \theta)} - b e^{-i(q_x x + \theta)}] &   \text{if }~~ |x|<d  \\ s te^{i(k_x x +\phi)} & \text{if }~~ x > d \end{array} \right .  \label{hole1} \eeq

\begin{figure}[t] 
\begin{center}
\includegraphics[width=90 mm]{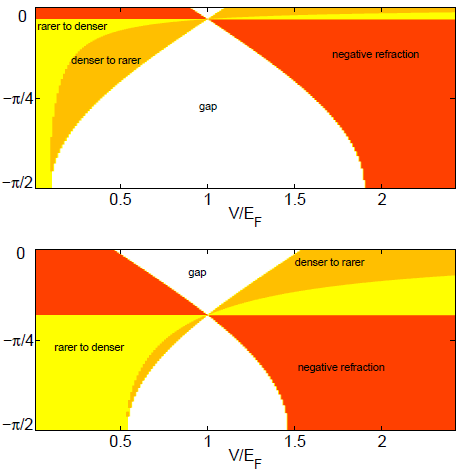}
\end{center}
\caption{Phase diagram of refraction angles $\theta$ as a function of $V$ on x-axis and $\phi$ on y-axis, and for an MVP barrier of width $D = 100 nm$ at $B = 0.1$ and $3T$. The gap corresponds to the region where TIR occurs.}
\label{refmap}
\end{figure}

These are similar in form to those for a pure magnetic barrier or an electrostatic step potential \ref{solutionelec}. But 
$\left(k_x,k_y\right)$ and $\left(q_x, k_y + \frac{1}{l_B}\right)$, namely the $x$ and $y$ components of the wavevector inside and outside the barrier regime are different. Here, $s,s'$ are $sgn(E-V)$ in the respective regions. Upon setting incident energy $E_F$ as $\hbar v_F k_F$, 
substitution of  Eqns.(\ref{particle1}) and (\ref{hole1}) in Eq.(\ref{decopemvp}) leads to 
\bea k_x^2 + k_y^2 & = & k_F^2, ~\text{with}~k_F=\frac{E_F}{\hbar v_F},
 ~ |x| > d \nonumber \\
q_x^2 + (k_y +\frac{1}{l_B})^2 & = & (k_F - \frac{V}{\hbar v_F})^2 
=k_F'^2, ~~ |x| \le d 
\label{energyemvp}\eea
The incidence angle $\phi$ and the refraction angle $\theta$ are given by  $\tan^{-1}(\frac{k_y}{k_x})$  and $\tan^{-1}(\frac{k_yl_B + 1}{q_x\ l_B})$ respectively. 
Eq.(\ref{energyemvp}) can then be rewritten to obtain the Snell's law analogue for electron waves of such massless Dirac fermions incident on the EMVP barrier as
\bea \sin |\theta| & = & 
S_F( \sin |\phi| + \text{sgn}(\phi)\frac{1}{k_F l_B}) \label{analogy11}  \\
& = & S_F \sin |\theta||_{V=0} \nonumber \\
S_F & = & \frac{k_F}{k_F'}=s'|\frac{E_F}{E_F-V}| ; ~ s'=sgn(E_F-V)  \nonumber \eea

Comparison between Eq.(\ref{analogy}) and Eq.(\ref{analogy11})  shows that 
the potential barrier effectively scales the refraction angle by the scale factor $S_F$ defined above. $S_F$ is a non-monotonic and discontinuous function of $V$ [Fig.\ref{raydiag}(a)] and we shall study its impact on the refraction of the incident electron wave. There are basically two regimes to be analysed depending upon the strength of electrostatic potential applied (i) $V<E_F$, (ii) $V>E_F$. For positive incidence angles $\phi  = (0,\pi/2)$, in the absence of any electrostatic potential the refraction angle $|\theta|$ is larger than the incidence angle $|\phi|$, thus the electrons are seen as passing from denser to rarer medium. As V increases from $0$ to $E_F$, the function $S_F$ and hence $\theta$ increases continuously (this is shown in fig.\ref{raydiag}(a) and (b) respectively) and eventually the electron wave suffers TIR as soon as the r.h.s of Eq. \ref{analogy11} becomes greater than 1. Thus, by increasing V it is possible to totally reflect an electron wave for any given $\phi$ and $B$.  Since the reflectivity of a magnetic barrier increases with higher $B$, this implies that the addition of  $V$ can effectively convert a weaker magnetic barrier into a stronger one. This effective control of the strength of the magnetic barrier by an electrostatic potential ( gate voltage) is a consequence of the ultrarelativistic nature of the charge carriers in graphene. In a subsequent section we shall provide a more formal reasoning for this feature. 

When V surpasses $E_F$, the sign of $\sin|\theta|$ will be opposite to the sign of $\sin|\phi|$. At smaller V close to $E_F$ there is still TIR, whereas for very high V much above $E_F$ the refraction becomes negative and the wave again retraces its path back in the barrier regime. This situation is depicted in Fig. \ref{raydiag}(b), where the wavevector for the refracted ray is shown changing with increasing V. A similar analysis can be done for negative incidence angles. We summarize this discussion by plotting $\sin|\theta|$ as a function of the incidence angle $\phi$ for different V and B values in Fig. \ref{refmap}. As can be seen, each of Fig.\ref{refmap}(a) and (b) are separated into an upper and a lower part by the line $\phi_s = \sin^{-1}(-\frac{1}{k_F l_B})$. In the lower part where $|\phi|$ is larger than $|\phi_s|$, with increasing $V$ the absolute value of the refracted angle, $|\theta|$ exhibits four phase regions: rarer $\to$ denser, denser $\to$ rarer, TIR gap, negative refraction. In the upper part, the four phase regions exhibit a different order with increasing $V$:  negative refraction, TIR gap, denser $\to$ rarer, rarer  $\to$ denser. All these different regions meet at the limiting point $(E_F, \phi_s)$, where the behaviour is singular and is discussed later on in this article in more detail.

\begin{figure}[t]
\begin{center}
\includegraphics[width=100 mm]{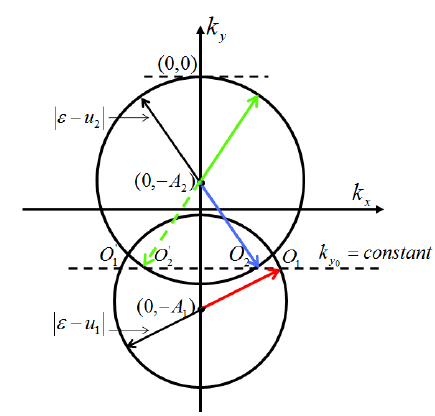}
\end{center}
\caption{Here $k_x$ and $k_y$ represent the transverse and
longitudinal components of the wave vector. The circles represent
solutions of the dispersion relations in the domains 1 and 2
and correspond to propagating waves.\cite{nori}}
\label{norifiglabel}
\end{figure}

The refraction laws for the electron waves incident on an MVP/EMVP barrier can be alternatively understood with the following geometric consideration\cite{nori}.
We consider two domains with scalar and vector potential $(V_1,A_1)$ and $(V_2,A_2)$ respectively. The electron wave being incident from domain 1 at an angle $\phi$ and refracted in domain 2 at angle $\theta$. Then the parallel and perpendicular wavevector components in domain 1 and 2 are related by the following dimensionless dispersion relations:

\beq k_{x_{1,2}}^2+(k_{y_{1,2}}+\mathcal{A}_{1,2})^2 = (\epsilon-u_{1,2})^2 \label{dispdomain1}\eeq

Each of the eq. \ref{dispdomain1} represents a circle in the $(k_x,k_y)$ plane, so that the wavevector for the propagating solutions will lie on the circle itself. Now, the refraction laws can be easily derived from fig.\ref{norifiglabel} as follows. Due to translational invariance along the y-direction, the perpendicular wavevector component takes the same value in both domains $k_{y_1} = k_{y_2} = k_{y_0}$. For a given $k_{y_0}$, propagating solution $k_x$ in domain 1 and 2 will take the value which lie at the intersection of line $k_{y_0} = \text{constant}$ with the corresponding circles. There are two such intersections for each of the circle. (This is shown in figure at $O_{1,2}$ and $O_{1,2}^{\prime}$). The physically meaningful intersection will be determined by the sign of  $(\epsilon-u_{1,2})$. For definiteness we assume the incident current density to be positive i.e. $k_{x_1}>0$. Then using the parametrization $k_{x_1} = (\epsilon-u_1)\cos\phi$, $\cos\phi$ will be positive or negative when $(\epsilon-u_1)> or < 0$ respectively. This means the incident wave will be directed along the vector joining the centre of circle with $O_1$ when $(\epsilon-u_1)>0 $ or  opposite to the vector joining the centre of circle with $O_1^{\prime}$ when $(\epsilon-u_1)< 0$. Similarly, using the parametrization $k_{x_2} = (\epsilon-u_2)\cos\theta$, the refracted current (which should possess the same sign as the incident current density) will be directed along the radius vector from the centre of the circle to the intersection point $O_2$ when $(\epsilon-u_2)>0$, or opposite to the radius vector from the centre of the circle to $O_2^{\prime}$ when $ (\epsilon-u_2)<0$.\
Simple geometry from the fig.\ref{norifiglabel} gives the following relation that defines the connection between the incident $\phi$ and refracted $\theta$ angles
\beq k_{y_0} = -\mathcal{A}_1-(\epsilon-u_1)\sin\phi =  -\mathcal{A}_2+(\epsilon-u_2)\sin\theta \nonumber \eeq
This relation gives the same condition as in eq. \ref{analogy11} for EMVP barrier.


\subsubsection{EMVP barrier: Transmission probability}

\begin{figure}[ht]
\begin{center}
\includegraphics[width=100 mm]{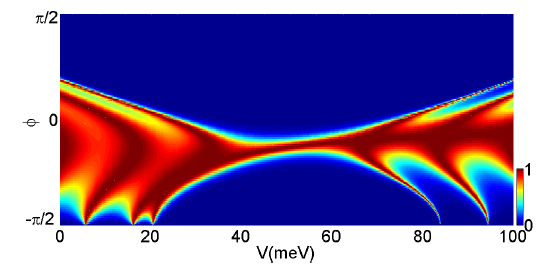}
\end{center}
\caption{Transmission in the presence of EMVP barrier in monolayer graphene, $B=1T$, $E_F = 50 meV$, $D = 100nm$.}
\label{Temvp}
\end{figure}

Transmission through a magnetic barrier gets strongly affected due to the presence of $V$. Again from the continuity at $x=\pm d$  in Eqs.(\ref{particle1}) and (\ref{hole1}), the transmission coefficient can be obtained as 
\beq t = \frac{2ss'e^{-ik_{x}D}\cos\phi\cos\theta}{ss'[e^{-iq_xD}\cos(\phi+\theta)+e^{iq_x D}\cos(\phi-\theta)]-2i\sin{q_xD}}, \label{temvp} \eeq

where $D = 2d$ and $q_x = k_F^{\prime}\cos\theta$. Here s and $s^{\prime}$ corresponds to sgn(E) and sgn(E-V) respectively. The expression for $t$ is same as the one for electrostatic potential barrier \ref{transformula} or MVP barrier. However the difference in the result is due to the change in the  expression for $q_{x}$, namely the $x$-component of the wave vector inside the barrier regime. In fig.\ref{Temvp} is plotted transmittance $T = t^{*}t$ as a function of $\phi$ and $V$ in the presence of an EMVP barrier. When the potential is lesser than but close to Fermi energy, transmission takes place over a very small window located asymmetrically along the $\phi$ axis. As the point $V=E_F$ is crossed, the function $S_F$ changes sign thereby changing the sign of refraction angle. Thus, the potential barrier induces more asymmetry in the transmission as against pure MVP barrier, this is clearly depicted in Fig. \ref{Temvp}. \

The point $V=E_{F}$ represents a singularity in the spectrum and demands a separate discussion. In the absence of magnetic barriers, such a point represents the zero modes for Dirac operators and leads to the emergence of new Dirac points. This has been discussed in a number of recent works considering Andreev Reflection in a graphene based NIS junction \cite{Shubhro} and for electronic states of graphene in a periodic potential \cite{Brey, CHPark2, CHPark3}. The presence of a magnetic barrier breaks the time reversal symmetry explicitly and these zero modes become the zero modes of the modified Dirac operator and the corresponding solutions are different. The equations satisfied by $\psi_{1,2}$ are 
\bea \left[\partial _x-(k_y+1/l_B)\right]\psi_1 &=& 0 \nonumber \\
\left[\partial _x+(k_y+1/l_B)\right]\psi_2 &=& 0 \nonumber \eea
for which the solutions are obtained as 
\bea \psi_1 &=& c_1e^{(k_y+1/l_B)x} \nonumber \\
\psi_2 &=& c_2e^{-(k_y+1/l_B)x} \nonumber \eea
By using the continuity condition at $x=\pm d$, the transmission expression is obtained to be 
\beq t = \frac{2\cos\phi e^{-ik_xD}}{e^{i\phi}e^{(k_y+1/l_B)D}+e^{-i\phi}e^{-(k_y+1/l_B)D}} \nonumber \eeq
The significance of such zero modes is that on the two sides of the singular point,
the relative sign between the $\phi$ and $\theta$ becomes opposite. 

\subsubsection{Conductance: Effect of various EMVP Barriers on Transport}

\begin{figure}[t]
\begin{center}
\includegraphics[width=100 mm]{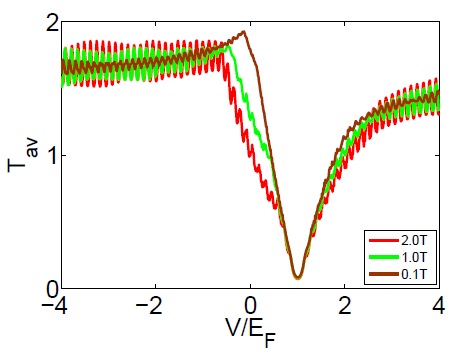}
\end{center}
\caption{Averaged transmission Vs voltage through a single EMVP barrier in monolayer graphene for different magnetic fields.}
\label{condlabel1}
\end{figure}
To see how the above angle-dependent transmission properties effect electron transport through such barriers, we plott the average transmission through the barrier as a function of the potential $V$ at various strengths of the magnetic barrier. The average transmission at a given  barrier strength $V$ and $B$ is defined as \cite{Masir1}
\beq  \langle T(B,V) \rangle = v_{F} \int_{-\frac{\pi}{2}}^{\frac{\pi}{2}} d\phi T(\phi, B,V) \cos \phi  \label{current} \eeq
This formula, when generalized to a range of energy levels, leads to the Landauer conductance $G \propto  2 \pi n e^2 \langle T(B,V) \rangle / h$. This has been plotted for three different strengths of magnetic barrier $B$ over a range of  potential barrier strength $V$  for a  single EMVP barrier in Fig. \ref{condlabel1}. The transmission shows a minimum as expected  when the point $qV=E_{F}$ is approached. The transmission grows on both side of this singular point and finally at higher $|V|$  oscillates around an average value.


\subsubsection{Generalisation to MVP/EMVP barriers of any arbitrary shape}
A general algorithm for the calculation of electron transmission in graphene through inhomogeneous electric and magnetic fields of any abitrary profile has been carried out in \cite{sameer} using transfer matrix method. The fields are invariant in one direction; and the method involves the division of the one-dimensional domain into slices and taking an appropriate approximation of potential form in each slice. The equation for each slice is solved and the continuity conditions are used at the interfaces of two such slices. The exact solution of the equation in each slice depends on the potential form chosen. It is shown that in the presence of piecewise constant scalar potential and piecewise linear vector potential the resulting equation admits parabolic cylindrical functions as the solution basis in moderate magnetic fields while the basis functions tend to complex exponentials in the presence of extremely small magnetic fields. Also it is shown that the localised charge distribution (electron and hole puddles) which arises due to the  presence of disorder in graphene \cite{yacobi} can be modelled as the charge as arising from a scalar potential distribution proportional to it. With this the transmission calculation corresponding to a one-dimensional potential extracted from the experimental data in ref. \cite{yacobi} is presented.
 The transport analysis with magnetic barriers in which the edges are smoothed out have also been carried out  \cite{milpas}- where a hyperbolic profile has been chosen. Here the corresponding Dirac equation can be analysed within the formalism of supersymmetric quantum mechanics, and leads to an exactly solvable model. Also, exact solutions have been given for a Dirac electron in the presence of an exponentially decaying magnetic field \cite{tkghosh}.

\subsection{Quantum Goos-H\"anchen Shift in single MVP and EMVP barriers} \label{GHsection}

Analogues of optical phenomena such as  refraction \cite{cheinov}, collimation \cite{parknanolett}, Fabry Perot interference \cite{peetersfabryperot}, Bragg reflection \cite{sgms} etc. in the ballistic transport regime for Dirac fermions in mono and bilayer graphene based structures in presence of scalar and vector potentials have been proposed. We discuss them in various sections of this review. 
In the following section 
 we discuss another important optical phenomenon, the Goos-H\"anchen (GH) effect \cite{GH, Newton}. The GH effect is a phenomenon of classical optics that describes the lateral shift between the centre of a reflected beam and that of incident beam when a total reflection occurs at the interface between two media. The shift occurs as the totally reflected ray undergoes a phase shift with respect to the incident beam, and this is is detectable since the extent of a real beam is always finite. The shift reverses sign if the second medium behaves like a metamaterial with negative refraction \cite{Berman, Shadridov}.  Such a lateral shift for totally as well as partially reflected  electron waves can also occur for non-relativistic electrons passing through a semiconductor barrier \cite{chen1}, magneto-electric semiconductor nanostructure \cite{chen2}. 
Recently, it has been shown  that ballistic electrons passing through a p-n interface in graphene \cite{BeenakerGH} also suffer a GH shift, which changes sign at certain angles of incidence. The analysis was extended for the case of MVP and EMVP barriers subsequently \cite{mssg}. To understand GH shift in graphene,  we consider the following wavepacket (beam) of electrons impinging on a p-n interface or a MVP/EMVP barrier at energy $E_F$:  
\beq \Psi_{in}(x,y) = \int_{-\infty}^{\infty} dk_{y} f(k_y-\bar{k}) e^{ik_y y + i k_{x}(k_y) x} \begin{bmatrix} 1 \\ e^{i \phi(k_{y}}) \end{bmatrix} 
\label{profilek} \eeq 
The envelope function ensures the  wavepacket is of finite size along the $y$-direction and is sharply peaked at $k_y=\bar{k}$. 
Thus, 
$\bar{k} \in ( 0,k_F)$ and the angle of incidence $\phi(\bar{k}_y) \in (0, \frac{\pi}{2})$. This fact is represented by writing the x-component of wavevector, $k_x$ as well as $\phi$ both as function of $k_y$ in  Eq.(\ref{profilek}). We take a finite beam with a  gaussian envelope such that 
\beq  f(k_y - \bar{k}) = \exp[ -\frac{(k_y -\bar{k})^2}{2\Delta_{k}^2}] \label{profile} \eeq 
When $\Delta_{k} \ll k_{F}$, we can approximate the $k_y$-dependent terms by a Taylor expansion around $\bar{k}$ and retaining only the first order term to get 
\beq \phi(k_{y}) \approx \phi(\bar{k}) + \frac {\partial \phi}{\partial k_{y}}|_{\bar{k}}(k_y-\bar{k});~ k_x(k_{y}) \approx k_{x}(\bar{k}) + 
\frac {\partial k_{x}}{\partial k_{y}}|_{\bar{k}}(k_y-\bar{k}) \eeq 
Substituting in Eq.(\ref{profilek}) and integrating, we obtain 
\beq \Psi_{in}  =  \sqrt{2 \pi \Delta_k^2} e^{i [\bar{k} y + k_x(\bar{k}) x]}\begin{bmatrix}
e^{-\frac{\Delta_k^2}{2}[y -\bar{y}_{+}^{in}]^2} \\ e^{-\frac{\Delta_k^2}{2}[y -\bar{y}_{-}^{in}]^2}e^{i \phi(\bar{k})} \end{bmatrix}, \eeq 
where 
\beq \bar{y}_{+}^{in} = -k'_x(\bar{k})x, \bar{y}_{-}^{in}= -k'_x(\bar{k})x - \phi'(\bar{k}) \eeq  
Here the primed $(\prime)$ quantities denote derivative taken wrt $k_y$.
Thus, upper and lower components of the spinorial wave function are localized at separate points along the $y$-axis.  

The reflected wavepacket can also be written in an analogous way by making the transformation $k_x$ to $-k_x$ and $\phi$ to $\pi - \phi$ as 
well as multiplying the reflection amplitude $r(k_y)=|r(k_y)|e^{i \phi_{r}(k_y)}$. 
The reflected wave is then 
\bea \Psi_{r}(x,y) & = & \int_{-\infty}^{\infty} dk_{y} f(k_y-\bar{k}) e^{ik_y y -i k_{x}(k_y) x}  r(k_y)\begin{bmatrix} 1 \\ -se^{-i \phi(k_{y}}) \end{bmatrix} 
\label{profilekr} \eea
Here again, $s=sgn(E_F)$. The spatial profile of the reflected wave can be again obtained by first expanding all $k_y$ dependent quantities around $\bar{k}$ and retaining only the first order terms and then integrating in Eq.(\ref{profilekr}). This leads to 

\bea \Psi_{r} & = & \sqrt{2 \pi \Delta_k^2} e^{i [\bar{k} y - k_x(\bar{k}) x]} |r(\bar{k})|\begin{bmatrix}
e^{-\frac{\Delta_k^2}{2}[y -\bar{y}_{+}^{r}]^2} \\ -se^{-\frac{\Delta_k^2}{2}[y -\bar{y}_{-}^{r}]^2}e^{-i [\phi(\bar{k})-\phi_{r}'(\bar{k})]} \end{bmatrix} \eea                                                 

\begin{figure}[t]
\begin{center}
\includegraphics[width=100 mm]{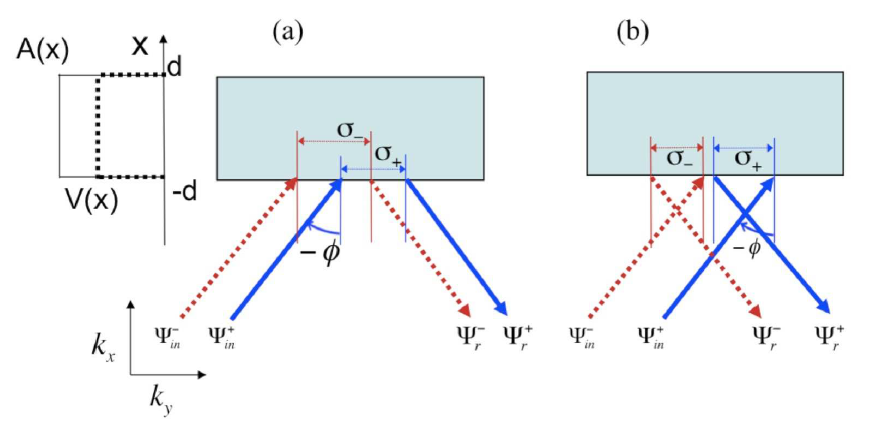}
\end{center}
\caption{GH shift for an EMVP barrier. The solid line corresponds to the upper component of the pseudospinor and the dotted line to the lower component. The GH shift can be either (a) positive or (b) negative.}
\label{figGHraylabel}
\end{figure}

 Here, $\bar{y}_{+}^{r}$ and $\bar{y}_{-}^{r}$ are given by \beq \bar{y}_{+}^{r} = -\phi'_{r}(\bar{k})+k'_x(\bar{k})x, \bar{y}_{-}^{r}= -\phi'_{r}(\bar{k}) +k'_x(\bar{k})x +\phi'(\bar{k}) \eeq 
The above expression shows that the upper as well as lower components get shifted because of  the phase factor. The GH shifts of the upper and lower components are respectively given by 
\bea \sigma_{+} & = & \bar{y}_{+}^{r} - \bar{y}_{+}^{in} = -\phi_{r}'(\bar{k}) + 2k_{x}'(\bar{k})x \nonumber \\
 \sigma_{-} & = & \bar{y}_{-}^{r} -\bar{y}_{-}^{in} = 2 \phi'(\bar{k}) -\phi_{r}'(\bar{k}) + 2k_{x}'(\bar{k})x \eea  
Thus, the average shift for an MVP or EMVP barrier  is 
\beq \sigma = \frac{1}{2} (\sigma_{+} + \sigma_{-}) = \phi'(\bar{k})-\phi_{r}'(\bar{k}) + 2k_{x}'(\bar{k})x \label{GHshift} \eeq 
The situation is depicted schematically in Fig. 
\ref{figGHraylabel}. The last term in the above expression is a coordinate dependent quantity and will get an equal and opposite contribution from the $-\phi_{r}'(\bar{k})$ term. The resultant $\sigma$ will thus be independent of the choice of the coordinate of the interface from which TIR will take place. 

\subsubsection{GH shift at p-n interface}

To calculate the GH shift occuring at the p-n interface, we first require to calculate the gradient of the phase of the refletion coefficient, the negative of which gives the GH shift. For this we match the wavefunctions at the interface $(x = -d)$ of the barrier such that the propagating wavefunction on the left of the interface should match with the evanescent wavefunction at the right of the interface.
 
\beq
\psi_1^{GH} = \left\{
\begin{array}{rl}
e^{ik_x x} + r e^{-ik_x x} & \text{if }~~ x < -d\\
a' e^{-\kappa (x+d)}  & \text{if }~~ x >-d  \\
\end{array} \right.
\label{particlegh}\eeq
\beq \psi_2^{GH} = \left\{
\begin{array}{rl}
s[e^{i (k_x x +  \phi)} - r e^{-i(k_x x + \phi)}] & \text{if }~~ x < -d\\
-i \gamma s' a'e^{-\kappa (x+d)}  &   \text{if }~~ x > -d 
\\ 
\end{array} \right . 
\label{holegh}
\eeq
Here $s = sgn(E_F),s'=sgn(E_F-V)$, and  
\beq  \gamma  =  \frac{\hbar v_F(\kappa + k_{y})}{E-V}, ~~ k_{y}^2  - \kappa^2  =  \left(\frac{E - V}{\hbar v_F}\right)^2 \nonumber \eeq
TIR at p-n interface will take place when $\kappa^2 = E_F^2 = (E_F-V)^2 > 0$, for which the incidence angle lies in the range $\phi>\phi_c = \sin^{-1}\left(\frac{V}{E_F}-1\right)$. The continuity of the wave function at $x=-d$ gives the reflection coefficient as 
\bea r & = & e^{-ik_x D} [\frac{ie^{i \phi} - ss'\gamma}{ie^{-i \phi} + ss' \gamma} ]=\exp(-ik_x D)\exp(2i\delta) \nonumber \\
\text{with} \tan \delta & = & \tan \phi + ss' \gamma \sec \phi \eea 
Clearly, the phase of the reflection coefficient is $ \phi_{r} = -k_x D + 2 \delta $. The GH shift can now be written as: 
\beq \sigma  =  \phi^{\prime}(\bar{k}) - 2 \delta'(\bar{k}),   ~~ \text{where} ~  \delta = \tan^{-1}(\tan\phi+ss^{\prime}\gamma\sec\phi) \label{ghfinal} \eeq

\begin{figure}[t]
\begin{center}
\includegraphics[width=100 mm]{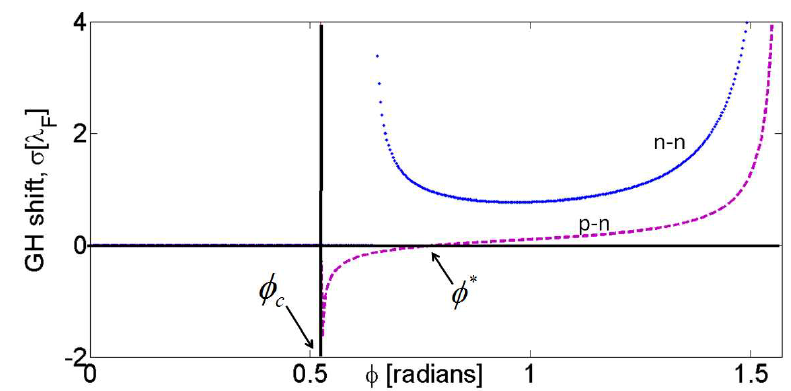}
\end{center}
\caption{Dependence on the angle of incidence $\phi$ of the GH
shift $\sigma$, for $U_0/E_F=1.5$ (dashed curve, p-n interface) and for $U_0/E_F=0.4$ (dotted curve, n-n interface).}
\label{GHpnfigure}
\end{figure}

Using this expression, one can calculate the GH shift at p-n interface.
\beq \sigma = \frac{\sin^2\phi+1-V/E_F}{\kappa\sin\phi\cos\phi} \label{ghatpninterface} \eeq
This is the same expression as the one obtained in \cite{BeenakerGH}. Since $\kappa>0$ for decaying evanescent solutions at the right of the p-n interface, for positive incidence angles the negative GH shift at the p-n interface $(E_F<V)$ will occur when $\sin^2\phi-(V/E_F-1)<0$. Thus, for angles of incidence $\phi_c<\phi<\phi^{*} = sin^{-1}\sqrt{\sin\phi_c}$, GH shift occurs in the backward direction.
This is explicitly shown in fig.\ref{GHpnfigure}.


\subsubsection{GH shift for MVP/EMVP barriers}
\begin{figure}[t]
\begin{center}
\includegraphics[width=120 mm]{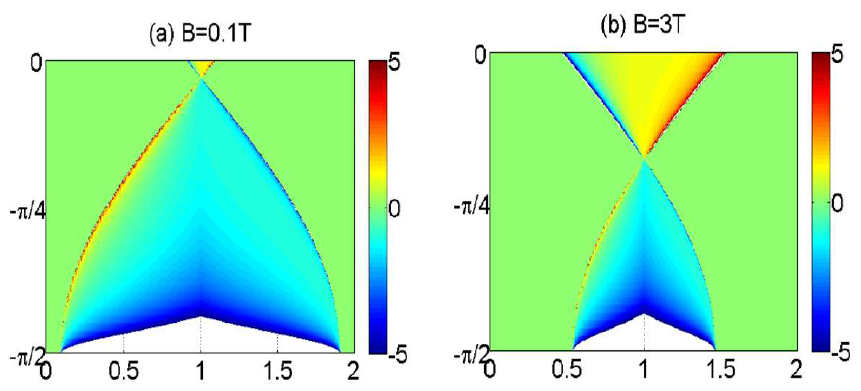}
\end{center}
\caption{\textit{(color online)} GH shift for an EMVP barrier at (a) $0.1$T and (b) $3$T. The $y$-axis corresponds to the incident angle $\phi$. The green regions have propagating solutions with no shift.}
\label{figGH}
\end{figure}

For pure MVP barrier, TIR will take place when $\left|sin\left|\theta\right|\right|  = sin\left|\phi\right| + sgn(\phi)\frac{1}{k_Fl_B} > 1$. This will happen when $\sin |\theta| > 1$
for  $0 < \phi \le  \frac{\pi}{2}$  and  when $\sin |\theta| < -1$  for  ${-\frac{\pi}{2}} \le \phi < 0$. In the latter case, this requires the wave vector to be negatively refracted at sufficiently high magnetic field before TIR occurs. This is already explained in section \ref{MVPsection} in detail.
The GH shift can then be obtained with the expression as in eq.\ref{ghfinal} with only 
\beq \gamma  =  \frac{\kappa + (k_{y} + \frac{1}{l_B})}{k_F}, 
~~ \left(k_{y} + \frac{1}{l_B}\right)^2  - \kappa^2  =  \left(\frac{E_F}{\hbar v_F}\right)^2  \nonumber \eeq

An EMVP barrier is essentially a p-n junction placed in a MVP barrier. 
For an EMVP barrier, TIR occurs for electrons incident from both sides of the surface normal but at different critical angles. For a given $\phi$, it is possible to change $V$ adiabatically and get the electron wave reflected over a range of $V$ satisfying $\left|sin\left|\theta\right|\right|> 1$. We can keep $\phi$ fixed and increase $V$. At $V=V_{c_1}$, TIR occurs when $\sin |\theta|  =  1$ giving
\beq  \frac{V_{c_1}}{E} = 1-[\sin|\phi| +sgn(\phi) \frac{1}{k_F l_B}] \nonumber \eeq
Upon further increasing $V$, the electron wave remains totally  
reflected till $V$ reaches the second critical value $V_{c_2}$ such that  $\sin |\theta| =  -1$ giving
\beq \frac{V_{c_2}}{E}  =  1 + [\sin |\phi| + sgn(\phi)\frac{1}{k_Fl_B}] \nonumber \eeq

TIR occurs in the range $V \in [V_{c_1} , V_{c_2}]$. In Fig. \ref{figGH} 
is plotted the GH shift over the entire range of $V$ and $\phi$.
At all other regions in the $\phi - V$ plane the GH shift is set to $0$ (green).
The explicit expression for the GH shift for an EMVP barrier in dimensionless form is 
\beq 
\sigma k_F  =  \frac{1 - 2 \frac{ k_F}{k_F'}\cos^2 \delta(1+ \tan \phi \tan \delta + ss' \frac{\gamma^2 +1}{\gamma^2 -1})} {\cos \phi}
\label{fullGH} \eeq
The $ \frac{\gamma^2 +1}{\gamma^2 -1}$ term in the numerator which is  $\coth \alpha$ or $\sec \beta$ 
diverges at the critical angle for TIR  and leads to a divergent GH shift. This is because just at the critical angle the wavevector lies in the interface of the two regions.
As a result in  Fig. \ref{figGH}, the border TIR region of finite GH shift shows the highest GH shift. 

The upper and lower parts of the curve are again separated by the line $\phi_s= \sin^{-1}\left(-\frac{1}{k_Fl_B}\right)$ . The left and right boundaries correspond to $V < E_F$ and $V > E_F$. In the lower part, the GH shift is mostly negative with the left and right boundaries having positive and negative refraction respectively.

Comparing Eq.(\ref{fullGH}) with Eq. (\ref{ghatpninterface}), we see that $\sigma(\phi) \ne - \sigma(-\phi)$. This is due to the non-specular nature of electron refraction at an EMVP barrier. This is an important difference in the quantum GH effect that occurs upon TIR by  an EMVP barrier as compared to TIR by a purely electrostatic barrier for which  $|\sigma(\phi)| = |\sigma(-\phi)|$ as can be seen from Eq.\ref{ghatpninterface}.


\subsubsection{ Lateral shift in the transmitted wave}

As explained in the previous section, Goos-H\"anchen shift is analogous to the phenomenon of the lateral shift of the light beam total internally reflected from dielectric surface. Chen et al. \cite{chen1,chen2} investigated the lateral shifts for Dirac fermions in transmission through monolayer graphene barrier, based on tunable transmission gap. This shift has same physical origin that is due to the beam reshaping since each plane wave component undergoes different phase shift, however it has nothing to do with the evanescent waves which plays an all important role in the lateral shift of total internally reflected wavefunction. For this reason this shift can be termed as Goos-H\"anchen like (GHL) shift and can be considered as an electronic analogue of the lateral shifts of light beam transmitted through a metamaterial slab.

According to stationary phase approximation \cite{stationaryphase} and on the same lines as explained for GH shift for the reflected beam in th previous section, the GHL shifts of the transmitted beam through a barrier in graphene can be obtained as \cite{chen1,chen2} $\sigma_t = -\left. \frac{\delta \phi_t}{\delta k_y}\right |_{k_{y_0}} $
where $ \phi_t$ refers to the phase of the transmission amplitude and the subscript corresponds to the wavevector at central incidence angle.
Using (\ref{transformula}), and assuming the form $t = |t|e^{i\phi_t}$, the transmission phase $\phi_t $ is obtained as
\beq \phi_t = tan^{-1}\left[\left( ss^{'}\sec\theta\sec\phi-\tan\theta\tan\phi\right)\tan q_xd\right] \nonumber\eeq
so that \beq \sigma_t =  -\frac{\delta \phi_t}{\delta k_y} = d\tan\phi_0\cdot T\cdot\left[ \left(2+\frac{k_0^2}{k_x^2}+\frac{k_0^2}{q_x^2}\right)\frac{\sin 2q_xd}{2q_xd}-\frac{k_0^2}{q_x^2} \right] \label{ghshiftsingleVbar} \eeq, 
where $k_0^2 = k_{y0}^2 - ss^{'}k_Fk_F^{'}$.
\begin{figure}
\begin{center}
\includegraphics[width=100 mm]{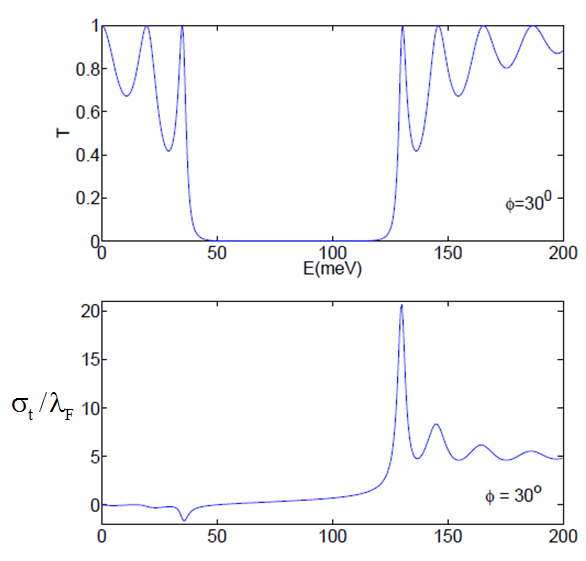}
\end{center}
\caption{Transmission gap (a) and GHL Shift (b) as a function of incident energy, E at $\phi = 30^{0}$, $U = 62meV$, $d=100nm$.}
\label{newgeoplots2}
\end{figure}

Since the product $ss^{'}$ appears in the GHL shift expression so the analysis of the expression can be classified into $ss^{'} = -1$ (Klein tunneling regime) and $ss^{'} = +1$ (classical motion regime).\


In the regime when $ss^{'} = -1$, $k_0^2 =  k_Fk_F^{'}+ k_{y0}^2 >0$ (for $ \phi>\phi_c $ and for all values of incident energy, E).
Then in accordance with (\ref{ghshiftsingleVbar}), the maximum absolute value GHL shift corresponds to  $q_xd = n\pi$ (These wavevector also corresponds to transmission resonances).
Hence \beq {\sigma_t}_{ q_xd = n\pi} = - \frac{k_0^2}{q_{x0}^2}d\tan\phi_0 \nonumber \eeq 
Thus in Klein tunneling regime, the GHL shifts are negative shifts when $\phi<\phi_c$.


In the other regime when $ss^{'} = 1$,  $k_0^2 =   k_{y0}^2-ss^{'}k_Fk_F^{'} = -k_0^{'2} <0$ (for $ \phi<\phi_c $ and for all values of incident energy, E).
Then in accordance with (\ref{ghshiftsingleVbar}), the maximum absolute value of GHL shift corresponds to  $q_xd = n\pi$ (This wavevector also corresponds to transmission resonances).
Hence \beq {\sigma_t}_{ q_xd = n\pi} = \frac{k_0^{'2}}{q_{x0}^2}d\tan\phi_0 \nonumber \eeq 
Thus in classical regime, the GHL shifts are always positive.

Thus we see that the GHL shifts in transmission can be negative or positive, and can be enhanced by the transmission resonances when the incidence angle is less than the critical angle for total reflection. This is clearly depicted in Fig.\ref{newgeoplots2}. 

\subsection{Fabry-Perot Interference in Graphene Heterojunctions: Effect of magnetic field}\label{FP}

\begin{figure}
\begin{center}
\includegraphics[width=100 mm]{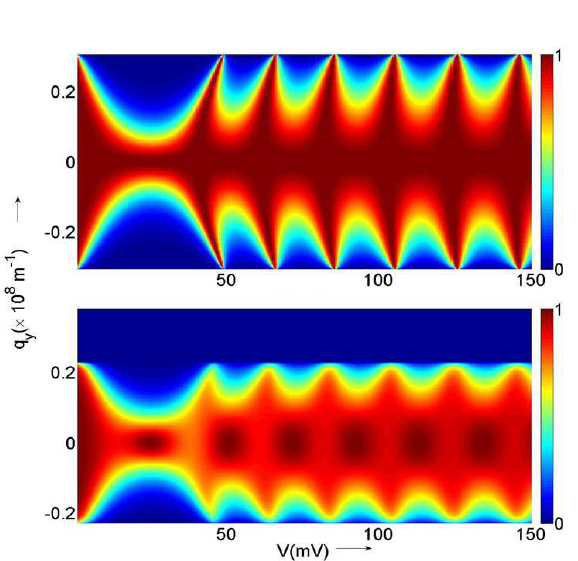}
\end{center} 
\caption{(\textit{Color online}) (a) Contour plot of the transmission through a single barrier Vs its height V and wave vector $k_y$ for D=100 nm. (a) $B=0$ (b) $B=0.1T$.}
\label{fabres}
\end{figure}
Another important optical analogue that can be studied in the ballistic transport regime of graphene is the formation 
of Fabry-Perot resonances. Study of such resonances is the linear and non-linear transport regime for non relativistic 
electrons for devices such as resonant tunnelling diodes is well known for a long time particularly after the pioneering work 
by \cite{Esaki, dutta}. A excellent review on the experimental progress on such resonances in  graphene based heterostructures in the ballistic transport regime can be found in \cite{Youngreview}. 

Below we revisit the analysis of potential barrier as a Fabry-Perot interferometer \cite{peetersfabryperot, shytov} and then we analyse that how the FP resonances get effected in the presence of magnetic field in graphene. \

\begin{figure}[t]
\begin{center}
\includegraphics[width=100 mm]{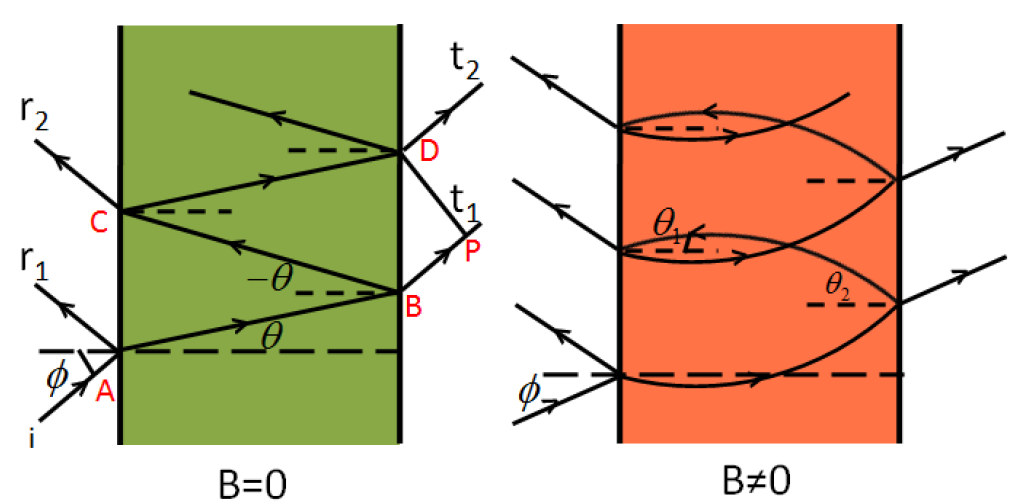}
\end{center}
\caption{Electron transmitting through a potential barrier in (a) absence (b) presence of magnetic field.}
\label{fabB}
\end{figure}

The two edges of the square potential barrier acts as the two interfaces of FP interferometer \cite{peetersfabryperot}, as shown in Fig. \ref{fabB}.
 As an electron wave is incident on the potential barrier at an angle $\phi$, it splits into the transmitted wave and reflected wave at the left edge $(x=-d)$. The transmitted wave propagate further to suffer another reflection(transmission) at the right edge $(x=+d)$. In this way it gets multiply reflected between the two edges of the barrier. \
The difference in the optical paths between the transmitted wave $t_1$ and $t_2$ which suffer consecutive reflections at the right interface is given as
\bea \Delta L &=& 2(1-V/E)(BC+CD)-BP \nonumber \\
 &=& \left(1-\frac{V}{E}\right)\frac{2D}{\cos\theta} - 2D\tan\theta\sin\phi \nonumber \\
 &=& \left(1-\frac{V}{E}\right)\frac{2D}{\cos\theta} - \frac{2D}{\cos\theta}\left(1-\frac{V}{E}\right)\sin^2\theta \nonumber \\
 &=& \left(1-\frac{V}{E}\right)2D\cos\theta ~~ \text{using} \sin\phi = \left(1-\frac{V}{E}\right)\sin\theta   \nonumber \eea
This means that the net phase difference between $t_1$ and $t_2$ is $ \delta = k_x\Delta L = 2q_xD  $.


Again we introduce $r$ and $t$ to be the reflection and transmission coefficient for the potential step outside the barrier, and $r^{\prime}$ and $t^{\prime}$ the corresponding coefficients for inside the barrier. In analogy with the optical waves, net transmission through the barrier is obtained as 
\beq  t_{tot} = tt^{\prime} + tt^{\prime}{r^{\prime}}^2e^{i\delta}+ . ~.~ . + tt^{\prime}{r^{\prime}}^{2(n-1)}e^{i(n-1)\delta} = tt^{\prime}/[1-{r^{\prime}}^{2}e^{i\delta}] \nonumber \eeq. 

Then, the total transmission probability $T = t_{tot}^{*}t_{tot}$ is obtained as
\beq T = 1/\left[1+\left(\frac{4|r^{\prime}|^2}{(1-|r^{\prime}|^2)^2}\right)\sin^2\delta/2 \right] \label{fabryT} \eeq

Proceeding with the calculation given in section 2, the reflection and transmission coefficients $r$, $r^{\prime}$, $t$ and $t^{\prime}$ can be obtained by matching the wavefunction at the interface of the potential step.
\bea \psi_{x<0} &=& \left( \begin{matrix} e^{ik_xx}+re^{-ik_xx} \\ s[e^{ik_xx+i\phi}-re^{-ik_xx-i\phi} ] \end{matrix}\right ) \nonumber \\
\psi_{x>0} &=& \left( \begin{matrix} t^{\prime}e^{iq_xx}\\ s^{\prime}t^{\prime}e^{iq_xx+i\theta} \end{matrix}\right) \nonumber \eea
Matching conditions at $x=0$ gives
\bea t = t^{\prime} &=& \frac{2\cos\phi}{s^{\prime}/s e^{i\theta}+e^{-i\phi}} \nonumber \\
r = r^{\prime} &=& \frac{1-s^{\prime}/s \cos(\phi-\theta)}{1+s^{\prime}/s \cos(\phi-\theta)} \label{stepcoeffs} \eea
Substituting \ref{stepcoeffs} in \ref{fabryT} we obtain the total transmission probability of a single barrier.
\beq T = \frac{\cos^2\theta\cos^2\phi}{(\cos q_xD\cos\theta\cos\phi)^2+\sin^2q_xD(1-ss^{\prime}\sin\theta\sin\phi)^2} \nonumber \eeq

Clearly the transmission resonances which are also the Fabry perot resonances occur at $q_xD = n\pi$. The energies at which these resonance occur are obtained as: 
\beq E = V \pm \hbar v_F[n^2\pi^2/D^2+k_y^2]^{1/2} \eeq
Also T is equal to 1 at normal incidence $\phi = 0$.\

To show the effect of magnetic field which is already mentioned in section\ref{KTsection} explicitly, we consider a potential barrier of height $V$ with a commensurate $\perp$ magnetic field $\bf{B}$ \cite{peetersfabryperot}
\bea V(x)=V, & \bs{B} = B\bs{\Theta}(d^2-x^2)\hat{z} & ~~ |x|<d \eea

With the similar analysis as in section \ref{finitemagbarsect}, the carriers in the presence of such potential form, obey the Schr$\ddot{o}$dinger like equation of the form:

\beq  \left [-\frac{\partial^2}{\partial\bar{x}^2}+\left (k_yl_B+\bar{x}\right )^2+1 \right ]\psi_{1,2} = \left(\epsilon-\tilde{v}\right)^2\psi_{1,2}.  ~~~ |x|<d \nonumber \eeq

Upon solving, the solutions can be obtained as:

\beq \psi_1(x) =  a D_{(\epsilon-\tilde{v})^2/2-1}(\sqrt{2}(\bar{x}+k_yl_B)) 
+ b D_{(\epsilon-\tilde{v})^2/2-1}(-\sqrt{2}(\bar{x}+k_yl_B))  \nonumber \eeq
\beq \psi_2(x) = a\frac{i\sqrt{2}}{(\epsilon-\tilde{v})} D_{(\epsilon-\tilde{v})^2/2}(\sqrt{2}(\bar{x}+k_yl_B)) 
+ b \frac{-i\sqrt{2}}{(\epsilon-\tilde{v})}D_{(\epsilon-\tilde{v})^2/2}(-\sqrt{2}(\bar{x}+k_yl_B)) \nonumber \eeq

The solutions in the outside regions namely $|x|>d$ remain same as in section \ref{finitemagbarsect}. By using the continuity condition to match the wavefunctions at the boundaries of the barrier we numerically evaluate the transmission as plotted in fig.\ref{fabres}. Since in the presence of magnetic field the y-component of the momentum varies inside the barrier as $k_y \rightarrow k_y-eBx/\hbar $, so the for the incidence angles at the two interfaces 1 and 2 to be of equal sign, $k_y(x_1)$ and $k_y(x_2)$ should have opposite signs ie. 
\bea k_y-eBx_1/\hbar = k_y+eBd/\hbar <0 \nonumber \\
k_y-eBx_2/\hbar = k_y-eB(d)/\hbar <0 \nonumber \eea
This gives the following condition on the y-component of momentum:
\beq -eBd/\hbar < k_y <eBd/\hbar  \nonumber \label{conditionkineticmomentum}\eeq
A fringe contrast from color minimum of transmission to the color maximum of the transmission on the two sides of the lines at $k_y = \pm eBd$ can be clearly seen in the Fig.\ref{fabres} which occurs as a consequence of the half a period shift in the transmission resonances. Occurance of such phase shift for charge carriers in graphene was originally studied for a harmonic potential in a uniform magnetic field by Shytov et al. \cite{shytov} and used as an experimental signature of Klein tunnelling \cite{kleinexp}. This has already been discussed in section \ref{KTsection}.
Transmission and conductance of massless Dirac fermions in the presence of multiple, disordered short range scatterers can also be understood in terms of resonant transport through double barrier structure. This has been discussed in detail in Ref. \cite{neetugreen}.


\section{Periodic lattice of MVP barriers}
Electron transport in graphene in presence of superlattices formed with electrostatic potential or magnetic barriers lead to large a body work in recent times\cite{periodic6,sgms,mssg,Martino,fractal1}. They lead to interesting physics and potent with the possibility of creating new devices 
in the ballistic transport regime since they significantly alter the band structure. On other hand they resemble closely the 
phenomena of light propagation through medium with periodically modulated refractive index/dielectric constant which has been studied in detail in optics literature and also led to many important applications \cite{yariv, kty}. In the following section we 
discuss this type of transport and their consequences. We shall also discuss the emergence of additional dirac points in these modified band structure and the resulting collimation of transport electrons using such superlattice structure. 
\subsection{Bandstructure modification}

In the previous section we have  shown how  MVP/EMVP barrier affects the ballistic transport in a profound way. Thus  it will be interesting to find out how a periodic arrangement of these barriers will effect the transport, what is the resulting band structure etc.. As mentioned earlier, it has a direct optical analogue, namely  with the electromagnetic propagation in periodic stratified media following the pioneering work by Yariv and collaborators \cite{yariv}. It also provides us relativistic version of ordinary Kronig-Penny 
model and its magnetic counterpart. 

Practically there will only be a finite number of barriers present in the system and the lattice translational symmetry will break down at the boundaries. However to simplify the analysis, we assume that the unit structure- a double EMVP barrier structure can be repeated infinitely along the $x$-axis. The magnetic field creating such a barrier $\bs{B}=B_z(x)\hat{z}=B\l_B[\delta(x+d)+ \delta(x-d) -2 B \delta(x)]\hat{z}$ together with a superimposed splitgate voltage geometry, hence forth will be called DEMVP barrier. 
Since changing or reversing $V$ can locally convert a charge-neutral region into a p-n or n-p junction, such a periodic barrier can also be thought of as a semiconductor heterostructure. It should be emphasized that this can lead to new device structures due to combined effect of the highly inhomogeneous and periodic magnetic fields and controllable voltages. 
The vector and scalar potentials are characterized by 
\bea  V(x) & = & -V; A_y(x) =B\ell_B, -d  < x < 0:~\text{region I} \nonumber \\
      V(x) & = & V ;   A_y(x) =-B \ell_B, 0 < x < d:~\text{region II}~\label{split-gate}  \eea

The schematic diagram in Fig. \ref{ferromag}(b) shows how such DEMVP barriers can be constructed by placing metal electrodes and insulating layers between PMA materials with alternating magnetizations and the graphene sheet. 

We consider each unit cell of size $D=2d$ for the MVP as well as for the electrostatic potential barriers. 
Thus, the n-th cell is given by the region between  $(n-1)D$  and $nD$. In the $\alpha$-th part of a unit cell, the wavefunction is
\bea \psi_1 & = & a_{n}^{\alpha} e^{i q^{n}_{\alpha x} (x - nD)} + b_{n}^{\alpha}e^{-i q^{n}_{\alpha x} (x - nD)} \\
\psi_2 & = & s_{n}^{\alpha} \left[a_{n}^{\alpha} e^{i [q^{n}_{\alpha x} (x - nD)+ \theta_{\alpha}]} - b_{n}^{\alpha}e^{-i [q^{n}_{\alpha x} (x - nD)+ \theta_{\alpha}] } \right] \nonumber \label{periodicsol} \eea
Here, 
\bea \alpha  =  1,2; a_{n}^{1}  =  a_n, b_{n}^{1}  =  b_n, & & 
 a_{n}^{2}  =  c_n, b_{n}^{2}  =  d_n; \nonumber \\
s_n^{1,2}=s_{1,2}; s_{1,2}=\text{sgn}(E　\pm V) & & 
q^{n}_{1x,2x} =  q_{1,2} \eea 
 The wavevectors $q_{(1,2)}$  are given as
\bea  q_1^2 + [k_y + \frac{1}{l_B}]^2 &=& [\frac{E_F+V}{\hbar v_F}]^2 \nonumber \\
q_2^2 + [k_y - \frac{1}{l_B}]^2 &=& [\frac{E_F-V}{\hbar v_F}]^2 \nonumber \eea

The exponential factor $e^{-nD}$ reveals the existence of lattice translational symmetry, which is not present for the isolated EMVP and DEMVP barriers.

The continuity of the wavefunction at the first interface at $x=(n-1)D$ gives 
\beq \begin{bmatrix}1 & 1 \\ s_{2}e^{i \theta_2} & - s_2e^{-i \theta_2} \end{bmatrix} \begin{bmatrix} c_{n-1} \\ d_{n-1} \end{bmatrix} 
= \begin{bmatrix} e^{-i q_1 D}  & b_n e^{i q_1 D}  \\ s_1 e^{i[q_1 D - \theta_1]} & - s_1 e^{i[q_1 D - \theta_1]}] \end{bmatrix} \begin{bmatrix} a_n \\ b_n \end{bmatrix}  \label{conti1} \eeq

Similarly, the continuity at the second interface at $x=(n-1)D+ d$ gives 
\beq \begin{bmatrix} e^{-iq_1\frac{D}{2}}  & e^{i q_1 \frac{D}{2}} \\
s_{1}  e^{-i[ q_1 \frac{D}{2}- \theta_1}]  & -s_1 e^{i[ q_1 \frac{D}{2}- \theta_1]}] \end{bmatrix} \begin{bmatrix}a_n \\ b_n \end{bmatrix}
 =  \begin{bmatrix} e^{-i q_2 \frac{D}{2}} &   e^{i q_1 \frac{D}{2}} \\ s_2[ e^{-i[ q_2 \frac{D}{2}- \theta_2]}]  & - s_2e^{i[ q_2 \frac{D}{2}-\theta_2]}] \end{bmatrix} 
\begin{bmatrix}c_n \\ d_n \end{bmatrix} \label{conti2} \eeq

Introducing a shorthand notations like,
\bea M_A = \begin{bmatrix}e^{-ik_x} & 0 \\ 0 & e^{ik_x}\end{bmatrix}; &  M_{B_{1,2}} = \begin{bmatrix}  e^{-iq_{1x,2x}d} & 0 \\ 0 &  e^{iq_{1x,2x}d} \end{bmatrix} \nonumber \\
M_{s,s_{1,2}} = \begin{bmatrix} 1 & 0 \\ 0 & s,s_{1,2} \end{bmatrix}; & M_{\theta_{1,2}} = \begin{bmatrix}1 & 1 \\ e^{i\theta_{1,2}} & -e^{-i\theta_{1,2}}  \end{bmatrix}\nonumber \label{definematrix}\eea

Now Eqns.(\ref{conti1}) and (\ref{conti2}) can be rewritten as 
\bea M_{s_{{2},{n-1}}}M_{\theta_2}\begin{bmatrix} c_{n-1} \\
d_{n-1} \end{bmatrix} &= & M_{s_{1,n}}M_{\theta_1} {M_{B_1}}^2\begin{bmatrix} a_{n} \\
b_{n} \end{bmatrix} \nonumber \\
M_{s_{1,n}}M_{\theta_1}M_{B_1}\begin{bmatrix} a_{n} \\ b_{n} \end{bmatrix} & = &
M_{s_{2,n}}{M_{\theta_2}}{M_{B_2}}\begin{bmatrix} c_{n} \\ d_{n} \end{bmatrix} \label{latticebc}\eea

\begin{figure}[t]
\includegraphics[width=150 mm]{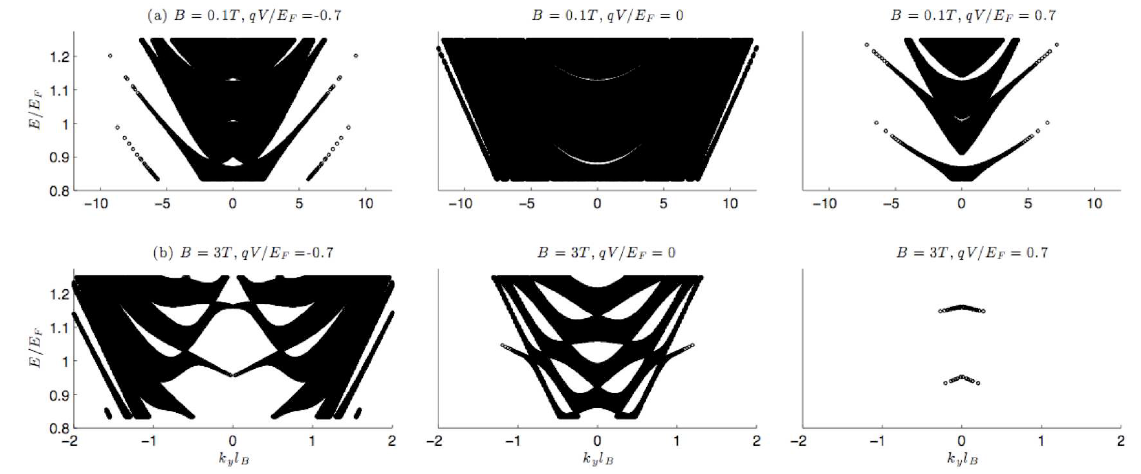}
\caption{{Band structure at some characteristic values of $V$ with 
$B$-field strength strengths (a) $3$ Tesla, and (b) $0.1$ Tesla.}}
\label{bsemvp}
\end{figure}

The above two matrix equations can be combined as 
\begin{widetext}
\beq \begin{bmatrix} c_{n-1} \\ d_{n-1} \end{bmatrix}  = 
M_{\theta_2}^{-1}M_{s_2}^{-1}M_{s_1}M_{\theta_1}M_{B_1}M_{\theta_1}^{-1}M_{s_1}^{-1}M_{s2}M_{\theta_2}M_{B_2}\begin{bmatrix} c_{n} \\ d_{n} \end{bmatrix}  =  \begin{bmatrix} K_{11} & K_{12} \\
K_{21} & K_{22} \end{bmatrix} \begin{bmatrix} c_{n} \\ d_{n} 
\end{bmatrix}\label{Blochcond1}\eeq
\end{widetext}
According to Bloch theorem,
\beq \begin{bmatrix} c_{n-1} \\ d_{n-1} \end{bmatrix} = e^{-iK D} \begin{bmatrix} c_{n} \\ d_{n} \end{bmatrix} \label{Blochcond2} \eeq 
The matrix $K_{mat}=\begin{bmatrix} K_{11} & K_{12} \\
K_{21} & K_{22} \end{bmatrix}$ is unimodular.  
From Eqs.(\ref{Blochcond1}) and (\ref{Blochcond2}) we obtain the eigenvalue equation 
\beq 
\begin{bmatrix} K_{11} & K_{12} \\
K_{21} & K_{22} \end{bmatrix} \begin{bmatrix} c_{n} \\ d_{n} \end{bmatrix}
 =e^{-iK D} \begin{bmatrix} c_{n-1} \\ d_{n-1} \end{bmatrix} \eeq
where $K$ is the Bloch momentum. 
The complex conjugate eigenvalues $\lambda$ are  given by  $det|K_{mat} - \lambda I|=0 $.This implies $\lambda_{1}+\lambda_{2}=\exp(-i K D)+\exp(i K D)$, which finally gives 
\beq K(\phi, B)=\frac{1}{2d} \cos^{-1}[\frac{1}{2}Tr(K_{mat})] \label{eqbandstructure} \eeq

Writing in terms of the wavevectors $q_1,q_2$ and the angles $\theta_1, \theta_2$, the above eigenvalue condition reads 
\beq \cos KD  =  \cos q_1 d \cos q_2 d + \sin q_1 d \sin q_2 d \times  \left[
\tan \theta_1 \tan \theta_2 - \frac{s_1s_2}{\cos \theta_1 \cos \theta_2}\right]
\label{band1} \eeq
This equation provides the band structure for a periodic DEMVP barrier superlattice in general for any $V \neq E$. By substituting $V=0$ in the above expression, the band structure for a periodic DMVP barrier superlattice can be obtained. This is an extension of the Kronig-Penney (KP) model to two-dimensional massless Dirac fermions. Thus, it is interesting to compare the DEMVP band structure  with other variants of the KP model. The original KP model describes Bloch waves in a one-dimensional periodic potential \cite{Kittel}. Several authors have also studied the relativistic KP model \cite{periodic1, periodic2, periodic4, periodic5}, where the motion considered is strictly one-dimensional. The non-relativistic KP model in periodic structures created by MVP barriers has also been studied \cite{peetersprl,periodic3}. For graphene charge carriers, such problems have also been studied for different types of periodic magnetic 
\cite{sgms,mssg, Martino, fractal1} as well as electrostatic \cite{periodic6} barriers. Magnetotransport studies in presence 
of periodic barriers \cite{Nasir} where the periodic modulation is sinusoidal was carried out for monolayer graphene and it 
was found that such periodic modulations Landau levels into bands which oscillate with the strength of the magnetic field.  

\begin{figure}[t]
\includegraphics[width=100 mm]{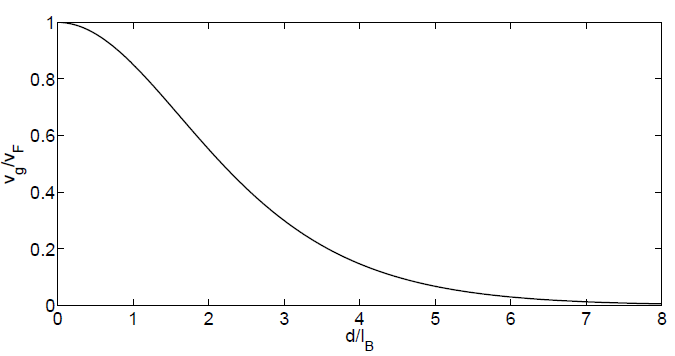}
\caption{{The group velocity $v_g$ (in units of $v_F$) at the neutrality point as a function of $d/l_B$.}}
\label{gpvel}
\end{figure}

In Fig. \ref{bsemvp} is plotted the band structure of corresponding to the periodic DEMVP barriers as a function of magnetic field $B$ and voltage $\frac{qV}{E_{F}}$. Conducting regions are seen over a wide range of $E$ with forbidden regions in between. At $0.1$T, the region near $V=0$ is mostly conducting for different values of energy where a forbidden region starts opening up  both to the left and right of the $V$ axis. The gaps are much wider than their counterparts for pure MVP barriers. Close to ${qV}\rightarrow E_F$, an extended forbidden zone appears. To understand more explicitly, the band structure is also plotted at $\frac{qV}{E_F}=0.7$ close to the singular point. 

Here, the conducting regions are intervened by large patches of forbidden zones. In comparison, at $\frac{qV}{E_F}=-0.7$, the behaviour is completely changed and the forbidden zone over the same range of $k_y l_B$ shrinks considerably (left column). For higher $V$ where $qV >E_F$, the system is conducting at almost all $E$. At $3$T and near zero $V$, a forbidden region opens up at various values of the energy and is much larger than at $0.1$T. The gapped regions to the left and right of zero $V$ at $3$T are located in a pronounced asymmetric manner as compared to when $B=0.1$T. This asymmetry in the band structure as a function of $V$ as well as the opening up of large forbidden zones for certain values of $V$ differentiates the transport through EMVP barriers from MVP barriers and thus, it provides more flexibility to tune such transport. \

As shown by various studies, carriers in graphene superlattices exhibit several interesting features \cite{parknat} \cite{Bliokh} \cite{parknanolett} that result from the particular electronic SL band structure. Below we briefly analyse the bandstructure of periodic DMVP barrier superlattice. 

\subsubsection{Dispersion in the vicinity of neutrality point}

\begin{figure}[t]
\begin{center}
\includegraphics[width=80 mm]{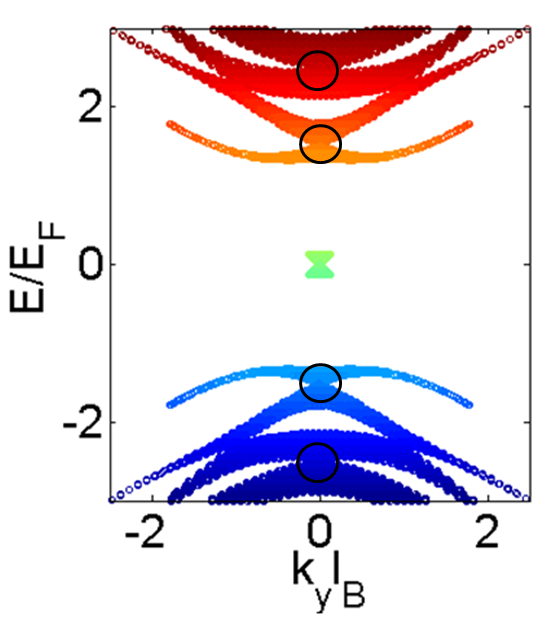}
\end{center} 
\caption{The circles emphasize the finite energy Dirac points at the values corresponding to Eq. \ref{extra}, for $B= 0.5T$, and $d=100nm$.}
\label{fdp}
\end{figure} 
We begin with analysing the dispersion relation in the vicinity of the neutrality point i.e.. close to zero energy. By rewriting the  Eq.\ref{band1} in the form of $f(E, k_x, k_y) = 0$, then expanding the $f$-function to the lowest order in $E$ and $k_y$ and then in $k_x$ in the vicinity of the neutrality point $(E=0, k_y=k_x=0)$, we obtain the dispersion relation:
\beq E = \pm \hbar v_Fv_0\sqrt{k_x^2+k_y^2}, ~~~ \text{with}~~  v_0 =\frac{d/l_B}{\sinh (d/l_B)} \label{neutralitypt}\eeq
Clearly, the above dispersion relation is isotropic $\left(\partial E/\partial k_x = \partial E/\partial k_y \right)$ with zero gap between the valence band and conduction band. Moreover the group velocity $v_g = v_0v_F$ depends on the width of the unit DMVP barrier as well as on the $B-$field strength. The group velocity is plotted in Fig. \ref{gpvel}, which clearly shows that $v_g$ is always smaller than the Fermi velocity and it monotonously decreases for increasing $d/l_B$, which is implied from the single MVP barrier analysis as well. 

\subsubsection{finite energy Dirac points}
Apart from the neutrality point described in Eq. \ref{neutralitypt}, many other degeneracy points can be seen in Fig.\ref{fdp} where the dispersion have a double cone like structure. The locations of these points occuring along the $k_y=0$ direction can be obtained analytically as 
\beq E_n = \pm \hbar v_F\sqrt{(1/l_B)^2+(n\pi/d)^2} \label{extra} \eeq 
Again by expnading the $f$-function in the vicinity of these points $E_n(k_x=0, k_y=0)$, the dispersion relation is obtained as 
\beq E = E_n\pm\hbar v_F\sqrt{v_{nx}^2k_x^2+v_{ny}^2k_y^2},~~~ n = 1,2,. . .  \label{extraDenergy} \eeq
with the group velocity components given by 
\bea v_{nx} &=& \frac{(n\pi)^2}{(d/l_B)^2+(n\pi)^2} \nonumber \\
v_{ny} &=& \frac{(d/l_B)^2}{(d/l_B)^2+(n\pi)^2} \nonumber \eea

\begin{figure}[t]
\begin{center}
\includegraphics[width=120 mm]{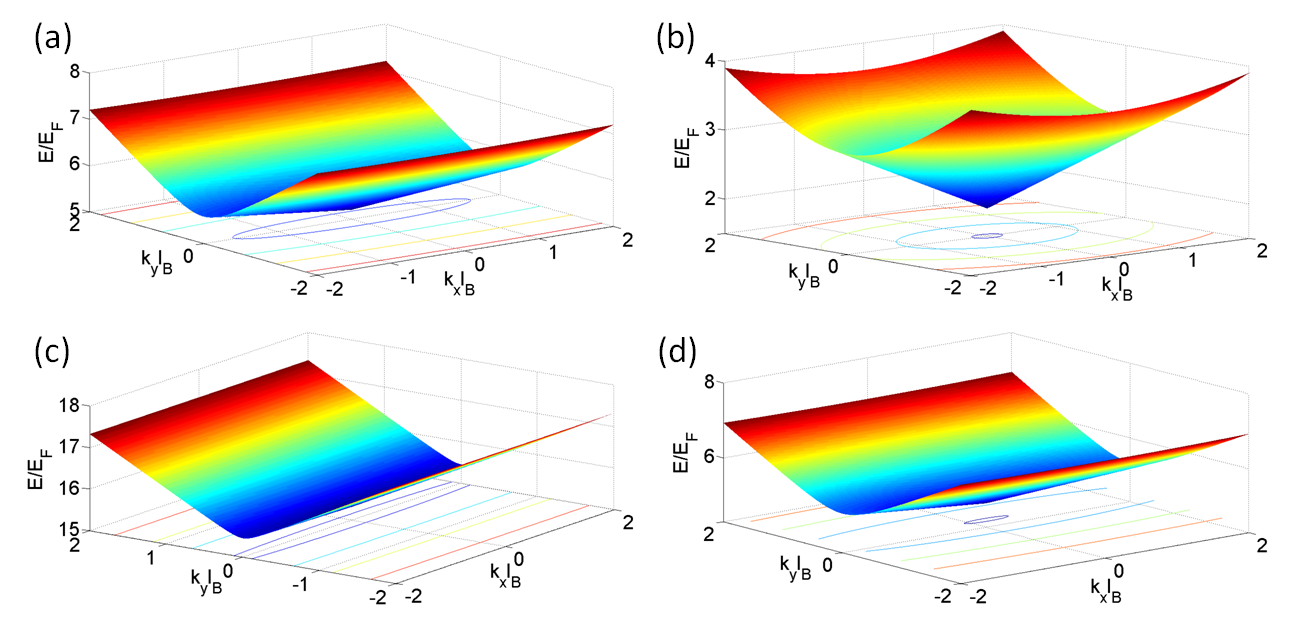}
\end{center}
 \caption{Plots of the dispersion relation at $n=1$ (top row) and $n=3$ (bottom row) for $B=0.1T$ (left column) and $B=1T$ (right column), $d=50nm$.}
\label{coll}
\end{figure}

Clearly the dispersion relation is linear in $k_x$ and $k_y$, which shows that in the presence of DMVP barrier superlattice, there are infinite number of degeneracy points close to which the dispersion has a double cone shape which is asymmetric in  $k_x$ and $k_y$ direction. These points are termed as "Extra Dirac points" in literature\cite{ana}\cite{nygen}.

In contrast to the dispersion relation at the neutrality point which is isotropic in nature, the dispersion at these finite energy extra Dirac points is anisotropic $\left(\partial E/\partial k_x \neq \partial E/\partial k_y \right)$.
As determined by Eq.\ref{extraDenergy}, we plot in Fig.\ref{coll}, the dispersion relation at $n=1, 3$ for different $B-$field strengths, $B=0.1,1T$. From this we infer a collimation along $k_y$ direction ie. $v_y \propto \partial E/\partial k_y \approx v_F$ and $v_x \approx 0$ as the contours become almost independent of $k_x$. This is similar to that found for a superlattice of electrostatic potential barriers \cite{parknanolett}.


\section{The transformation of Magnetic to Electric Field for dirac fermions in graphene: a case for relativistic invariance}
 For the electron transport in presence of a transverse uniform magnetic field it was 
shown by Lukose {\it et al.} \cite{lukose} that the Landau levels gets modified when the system is subjected to
an electric field, finally leading to the Landau level collapse through the renormalization of the cyclotron frequency.
A similar rescaling of the magnetic field strength took place when a suitable gate voltage is applied
 for the magnetic barrier problem as well \cite{mssg} as discussed in section \ref{EMVP}. To summarize one can change the strength of a magnetic field by suitably applying a 
get voltage. A more detailed analysis done later on \cite{tan} how the relativistic invariance of the effective equation 
that governs the charge carriers in graphene leads to this interconvertibility of electric and magnetic field. 
In this section we shall describe this approach. 

According to Special theory of Relativity, the laws of physics must be formed in a form so that it do not allow one to distinguish between the frames of reference which are moving with a constant relative velocity. The transformation between such frames are described by Lorentz transfromations. Like Maxwell's equations, Dirac equation is also Lorentz covariant.

So, for the charge carriers in graphene obeying "Dirac-like" equation, the Lorentz covariant structure of Hamlitonian can be used to find an analogy with special relativity. The analogy  extends to problems involving coupling of electrons to external electric or magnetic fields. It is known from from special Relativity that in the presence of electromagnetic fields there are two invariants; namely $E^2-B^2$ and $\bs{E.B}$ which remain unchanged in transition from one inertial frame to another \cite{landaulif}. This means, by means of Lorentz transformation we can always give $\bs{E}$ and $\bs{B}$ any arbitrary values, subject to only condition that  $E^2-B^2$ and $\bs{E.B}$ have fixed values. Particularly, if $\bs{E.B} = 0$, then we can always find a reference frame in which $\bs{E}=0$ or $\bs{B}=0$ (according as $E^2-B^2 < $ or $> 0)$, ie. the field is purely electric or purely magnetic. In the following discussion we'll first find the transformation laws for the graphene charge carriers and then we analyse the consequences of application of Lorentz boost \cite{lukose, tan,shytovguthesispaper}. For this, it will be convenient to write graphene equation in Lorentz - covariant form \cite{shytovguthesispaper}:
\beq \gamma^{\mu}\left(p_{\mu}-a_{\mu}\right)\Psi = 0 \label{covdirac} \eeq
where the $2\times 2$ $\gamma^{\mu}$ matrices satisfy the anticommutation relation $[\gamma^{\mu},\gamma^{\nu}]_{+} = g^{\mu\nu}$; $g^{\mu\nu}$ is the metric -tensor. We can explitly write the gamma-matrices as: $\gamma^{0} = \sigma_z $, $\gamma^{1} = -i\sigma_y$, $\gamma^{2} = -i\sigma_x$, where $\sigma_x,\sigma_y,\sigma_z$ are Pauli matrices. Here we use the space-time notation for coordinates as $x_{\mu} = (v_Ft,x_1,x_2)$, momenta $p_\mu = \hbar(iv_F^{-1}\partial_t, -i{\partial_x}_1,-i{\partial_x}_2)$, and external field $a_{\mu} = (a_0,a_1,a_2) = \left(-\frac{V}{v_F}, -\frac{eA_x}{c}, -\frac{eA_y}{c} \right)$.

The above Eq.\ref{covdirac} is invariant under Lorentz transformation. The transformation laws can be obtained on the same lines as for the general $(3+1)$ Dirac equation. The space-time coordinates, momenta and fields transform as:
\bea x^{\mu^{\prime}} &=&\Lambda^{\mu^{\prime}}_{\mu}x^{\mu},  ~~ p_{\mu^{\prime}} = \Lambda^{\mu^{\prime}}_{\mu}p_{\mu}, ~~  a_{\mu^{\prime}} = \Lambda^{\mu^{\prime}}_{\mu}a_{\mu} \\
\Psi^{\prime} &=& S(\Lambda)\psi \nonumber \eea
where $S(\Lambda) = exp\left(\frac{1}{8}\omega_{\mu\nu}[\gamma^{\mu},\gamma^{\nu}]\right)$ for $\Lambda = exp(\omega)$. 

We consider the magnetic field modulations where the field strengths vary in $x$ direction and are constant in y-direction and scalar potential V = V(x), then $a_1=0$ in Eq.\ref{covdirac}. Writing the solutions as $\Psi(x,y,t) = e^{ik_yy}e^{-iEt/\hbar}$, we obtain 
\beq E\psi(x) = \left[ -i\hbar v_F\sigma_x \partial_x + v_F\sigma_y\left(\hbar k_y+\frac{e}{c}A(x)\right)+V(x)\right]\psi(x) \label{parkeq1}\eeq

Now we apply two set of transformations \cite{tan} (A) a lorentz boost in y direction with rapidity $i\pi/2$, followed by (B) a reflection about the y-axis. These two transformations together make a complex Lorentz boost as $det(\Lambda_A)det(\Lambda_B) = 1$. 
With this complex Lorentz boost applied the energy and momentum transforms as :
\bea E &=& -i\hbar v_Fk_y^{\prime} \nonumber \\
k_y &=& iE^{\prime}/\hbar v_F \nonumber \eea

And transformed Eq. \ref{parkeq1} becomes
\beq E^{\prime}\psi^{\prime}(x) = \left[ -i\hbar v_F\sigma_x \partial_x + \sigma_y\left(\hbar v_F k_y^{\prime}-iV(x)\right)+\frac{e v_F}{c}iA(x)\right]\psi^{\prime}(x) \label{parkeq2}\eeq

On comparing \ref{parkeq1} and \ref{parkeq2} it can be clearly seen that it is the imaginary scalar potential which is now coupling with the y-component of momentum and the role of scalar potential is now played by the external vector potential (again  imaginary). 

This section concludes our reviewing various aspects of ballistic electron transport in monolayer graphene 
through magnetic barriers, tuned by gate voltage and the associated optical analogy. In the subsequent section we shall
discuss the electron transport in bilayer graphene in presence of such magnetic barriers.

\section{Magnetic barriers in bilayer graphene} \label{bilayer}
\subsection{Transmission through magnetic barriers}
In this section, we consider the effect of such inhomogenous magnetic fields for the case of unbiased bilayer graphene (BLG). \
Bilayer graphene is modelled as two coupled hexagonal lattices including inequivalent sites $a, b$ and $c, d$ in the bottom and top layers respectively. The present work discusses bilayer graphene in bernal stacking. In such an arrangement, the bottom and top layers  are shifted with respect to each other in such a manner that $c$ sites are exactly above the $a$ sites and $b$ sites are exactly in the middle of the hexagons of the bottom layer. In addition to intralayer interactions, there are interlayer interactions present in the system. The band structure is derived after taking into account two possible ways of $A\rightleftharpoons C$ hopping \cite{asymfalco}: via the dimer state or due to weak but direct $a-c$ coupling. There are other  weaker tunnelling processes \cite{dressdress} which are neglected in this tight-binding approximation which we consider. The tight-binding Hamiltonian \cite{asymfalco} derived in this way spans over the  basis  states that are located at $a, b, c, d$. The effective low energy Hamiltonian near the Fermi level \cite{asymfalco} obtained within the $\bs{k} \cdot \bs{p}$ approximation from this tight binding Hamiltonian in this case is different from the corresponding one  given in Eq.(\ref{graham}) for monolayer graphene.

The effective $2\times 2$ Hamiltonian is given as
\beq H_{eff} = -\frac{\hbar^2}{2m}\left(\begin{matrix} 0 & (k_x-ik_y)^2 \\ (k_x-ik_y)^2 & 0 \end{matrix} \right) \nonumber \eeq
 The solutions of these equations are massive Dirac fermions since they have a quadratic dispersion and non-zero effective mass m. The applicability of this model Hamiltonian to study transport through a scalar potential was discussed in detail in \cite{poole}. In the presence of scalar potential $V(x)$, under normal incidence $k_y=0$, pseudospin $\sigma_x$ is still conserved, but the Hamiltonian is like that of a Schr$\ddot{o}$dinger particle; i.e., $H_{eff} = -\frac{\hbar^2}{2m}k_x^2\sigma_x+V(x)I$. The probability amplitude of corresponding charge carriers exponentially decays like usual Schr$\ddot{o}$dinger particles inside the barrier leading to negligible transmission \cite{KSN3, todorovskiy, parkprb84}. Drawing an analogy with optical transmission we call this phenomenon Klein reflection. 

Below we calculate the effect of magnetic barrier for the case of bilayer graphene (BLG).
The electronic wavefunction depends on the spatial profile of the vector potential and the boundary conditions when an electron is incident from a region with $B=0$ to a region where $B \neq 0$. We show this by explicitly calculating transmission by a transfer matrix approach \cite{masir}, \cite{nilsson}.
Chiral charge carriers in BLG obey:
\beq  i\hbar\frac{\partial \Psi (x,y)}{\partial t} = H\Psi,~~~ H=\begin{bmatrix} V_1 & \Pi & \epsilon_t & 0 \\ \Pi^{+} &V_1 & 0 & 0 \\ \epsilon_t & 0 & V_2 & \Pi^{+} \\ 0 & 0 & \Pi & V_2 \end{bmatrix} \label{hammat} \eeq
Here,  they are described by a $4$-component spinor $\Psi(x,y)  =\begin{matrix} ( \Psi_a & \Psi_b & \Psi_c & \Psi_d)^T \end{matrix}$ and a $4 \times 4$  Hamiltonian in the presence of a magnetic barrier  and an electrostatic potential. $\Pi = v_F[p_x+i(p_y+eA/c)]$  with  $v_{F} =10^6$ m/s and $\epsilon_t$ is the tunnel coupling between the two layers. $ V_1$ and $V_2$ are the potentials at the two layers. 
The field profile taken here is $\bs{B}=B \Theta(x^2 - d^2) \hat{z}$. 
As pointed out earlier that such a sharp  profile is routinely created in magnetic recording media where bit lengths can be varied over  $10-100$nm range using suitable domain engineering \cite{terris, IEEE}.  Since this length scale is much larger than the A-A lattice constant of $0.246$nm in graphene but also much smaller than the Fermi wavelength  $\lambda_{F} =50$nm (at $E_{F}= 17$meV) of the incident electrons in bilayer graphene, the inhomogenous field pattern can be well approximated by the sharp edged profile proposed here. For bilayer graphene also, such sharp fields produced either by a patterned media layer or by a close-packed nanowire array can be easily integrated in the $\mathrm{SiO}_{2}$ underlayer present in typical top- and bottom-gated graphene devices for Klein tunnelling studies \cite{kleinexp}. The fields and voltages needed are within range of values already being used.

We are interested in the stationary state solutions of the form $\Psi = \psi(x,y)e^{-iEt/\hbar}$. Since $[H,p_y] = 0$, we can assume solutions like $\Psi_\alpha(x,y) = \phi_\alpha(x)e^{ik_y y}$, for $\alpha =  a,b,c,d $. Substituting this into Eq.\ref{hammat}, and using $l_B =\sqrt{\frac{\hbar c}{eB}}$ and $\epsilon_B = \frac{\hbar v_F}{l_B} $  as units of length scale and energy scale respectively, to define the dimensionless quantities $x\rightarrow\frac{x}{\l_B}$ , $v_{1,2}\rightarrow\frac{V_{1,2}}{\epsilon_B} $, $\epsilon^{\prime}_t\rightarrow\frac{\epsilon_t}{\epsilon_B}$ and $\epsilon \rightarrow \frac{E}{\epsilon_B}$, we obtain the following set of coupled equations:

\bea -i[ d/dx - (k_y+eA_yl_B/\hbar c)]\phi_b + \epsilon_t^{\prime}\phi_c = (\epsilon^{\prime}-\delta)\phi_a \nonumber \\
 -i[ d/dx +(k_y+eA_yl_B/\hbar c)]\phi_a = (\epsilon^{\prime}-\delta)\phi_b  \nonumber \\
 -i[ d/dx + (k_y+eA_yl_B/\hbar c)]\phi_d + \epsilon_t^{\prime}\phi_a = (\epsilon^{\prime}+\delta)\phi_c  \nonumber \\
 -i[ d/dx -(k_y+eA_yl_B/\hbar c)]\phi_c = (\epsilon^{\prime}+\delta)\phi_d  \label{set2} \eea

Here \beq
A_y = \left\{ \begin{array}{ll}
 sgn(x)Bd/2  & \textrm{if} |x|>\frac{d}{2}\\
Bx & \textrm{if} |x| < \frac{d}{2}
\end{array} \right.
\eeq
$sgn(x)$ represents sign of $x$, $\epsilon^{\prime}=\epsilon-v_0$ with  $v_0=(v_1+v_2)/2=\frac{V_{+}}{\epsilon_B}$ and $\delta = (v_1-v_2)/2 = \frac{\Delta}{\epsilon_B}$.

\begin{figure}
\begin{center}
\includegraphics[width=150 mm]{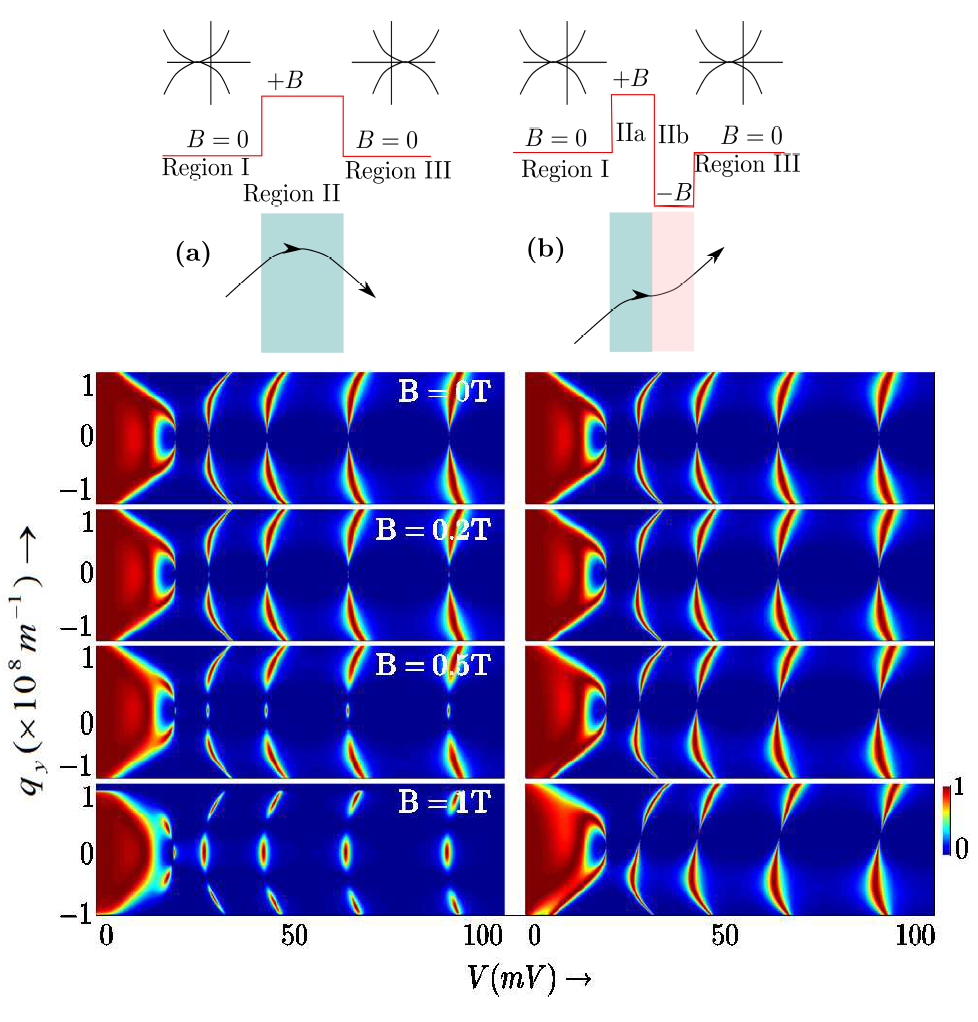}
\end{center}
\caption{Transmission, $T$ Vs $V$ and $q_y$ for various fields. (a) Barrier,  (b) Barrier+well. $\Delta=0$,  $d=50$nm, $E_F=17$meV.
}
\label{bilayerTr}
\end{figure} 

The solutions can be explicitly obtained  in the three regions $ x < -\frac{d}{2}$ (region I), $-\frac{d}{2} \le x \le \frac{d}{2}$ (region II) and $x > \frac{d}{2}$ (region III) as follows:

Region I($j=1$) and III($j=3$) are constant vector potential regions. We substitute $\phi_{\alpha}(x) =\phi_{\alpha}(0)e^{iq_x x}$ in Eq.\ref{set2}. With this, the dispersion relation becomes:
\beq [-q(x)^2 + (\epsilon^{\prime}+\delta)^2][-q(x)^2 + (\epsilon^{\prime}-\delta)^2]= ({\epsilon^{\prime}}^2-\delta^2){t^{\prime}}^2  \label{dispI} \eeq
Here, $q(x)^2 = q_{x}^2 + q_{y}^2$ with $q_{y}(x) = k_y + sgn(x) \pi \frac{\Phi}{\Phi_0}$.  Also, $\frac{\Phi}{\Phi_0}$ is the total magnetic flux through an area $d \l_B$ in units of the flux quantum $\Phi_0=\frac{hc}{e}$. The above Eq.\ref{dispI} leads to propagating as well as evanescent wave solutions. The unbiased BLG (which is the case considered here), corresponds to $\delta=0$, and $v_{1,2}=v_0=0$ inregion I and III. Then the complete wave function in region I and III can be given as:
\beq \Psi_{j}(x,y) =   \mathcal{M}_{0}(x)e^{ik_y y}(\begin{matrix} a_j & b_j & c_j & d_j )^T \end{matrix} \eeq

\beq
 \mathcal{M}_{0}(x)  =  \frac{1}{\epsilon'}\left[ \begin{matrix} {\epsilon'}e^{i q_x x} & {\epsilon'}e^{-i q_x x} & {\epsilon'}e^{-\kappa_x x} & {\epsilon'}e^{ \kappa_x x}  \\  
[{q_x-iq_{y}(x)}]e^{iq_x x} & -[{q_x+iq_{y}(x)}]e^{-i q_x x} &  i[{ \kappa_x-q_{y}(x)}]e^{- \kappa_x x}    &     -i[{\kappa_x+q_{y}(x)}]e^{\kappa_x x} \\ 
-{\epsilon'}e^{i q_x x} &-{\epsilon'} e^{-i q_x x} & {\epsilon'}e^{-\kappa_x x} & {\epsilon'}e^{ \kappa_x x}     \\  
- [{q_x+iq_{y}(x)}]e^{iq_x x} & [{q_x-iq_{y}(x)}]e^{-i q_x x} &  i[{ \kappa_x+q_{y}(x)}]e^{- \kappa_x x}    &     -i[{\kappa_x-q_{y}(x)}]e^{\kappa_x x}    \end{matrix} \right]  
\nonumber \eeq

For the wavefunction solutions in region II, where the magnetic field is finite, we introduce $(x+k_y) = z/\sqrt{2}$ and rewrite Eq.\ref{set2} as:

\bea -i\sqrt{2}[ d/dz - z/2]\phi_b + \epsilon_t^{\prime}\phi_c = (\epsilon^{\prime}-\delta) \phi_a  \nonumber \\
 -i\sqrt{2}[ d/dz +z/2]\phi_a = (\epsilon^{\prime}-\delta) \phi_b  \nonumber  \\
 -i\sqrt{2}[ d/dz + z/2]\phi_d + \epsilon_t^{\prime}\phi_a = (\epsilon^{\prime}+\delta) \phi_c  \nonumber  \\ 
 -i\sqrt{2}[ d/dz - z/2]\phi_c = (\epsilon^{\prime}+\delta) \phi_d  \label{mageq} \eea 

With this, we obtain two coupled equations in $\phi_a$ and $\phi_c$ which looks like 
\bea \left( (\epsilon^{\prime}+\delta)^2+2(d^2/dz^2-z^2/4-1/2) \right )\phi_c=(\epsilon^{\prime}+\delta) \epsilon_t^{\prime}\phi_a  \nonumber \\
 \left( (\epsilon^{\prime}-\delta)^2+2(d^2/dz^2-z^2/4+1/2) \right ) \phi_a= (\epsilon^{\prime}-\delta)\epsilon_t^{\prime}\phi_c  \eea

The above two equations can be solved to obtain a fourth order differential equation which can be decomposed in two second order differential equations.
For $\phi_a$, we obtain : 
\bea (d^2/dz^2-z^2/4+\gamma_{+}/2)(d^2/dz^2-z^2/4+\gamma_{-}/2)\phi_a= 0, \nonumber \\
\gamma_\pm = {\epsilon^{\prime}}^{2} + \delta^2 \pm [{(1-2\delta\epsilon^{\prime})}^2+{(\epsilon^{\prime}}^{2} - \delta^2){\epsilon_t^{\prime}}^2]^{1/2} \eea      

The above equation admits the solutions which are of the form of parabolic cylindrical functions \cite{gradshteyn}. For the case of unbiased BLG the complete wavefunction solution in region II can be given in terms of a compact matrix as:

\beq \mathcal{M}_{B}(x)  =  \begin{bmatrix} D_{p^{+}}(z) & D_{p^{-}}(z) & D_{p^{+}}(-z) & D_{p^{-}}(-z) \\
 \varepsilon_{1}^{*}p^{+}D_{p^{+}-1}( z) & \varepsilon_{1}^{*}p^{-} D_{p^{-}-1}( z) & \varepsilon_{1}p^{+}D_{p^{+}-1}( -z) &  \varepsilon_{1}p^{-}D_{p^{-}-1}( -z) \\ \varepsilon_{2}^{+}D_{p^{+}}( z) & \varepsilon_{2}^{-}D_{p^{-}}( z) & \varepsilon_{2}^{+}D_{p^{+}}( -z)& \varepsilon_{2}^{-}D_{p^{-}}( -z)\\ \
\varepsilon_{1}\varepsilon_{2}^{+}D_{p^{+}+1}(z)& \varepsilon_{1}\varepsilon_{2}^{-}
D_{p^{-}+1}(z)& \varepsilon_{1}^{*}\varepsilon_{2}^{+}D_{p^{+}+1}(-z)&
\varepsilon_{1}^{*}\varepsilon_{2}^{-}D_{p^{-}+1}(-z) \end{bmatrix}  \nonumber \eeq

Here D corresponds to parabolic cylindrical function of argument z and order $p^{\pm} = (\gamma_{\pm}-1)/2$. Also $\varepsilon_{1}=\frac{i \sqrt{2}}{\epsilon^{\prime}}$ and $\varepsilon_{2}^{\pm}= \frac{\epsilon^{\prime}}{\epsilon^{\prime}_t}-\frac{2 p^{\pm}}{\epsilon_t^{\prime}\epsilon^{\prime}}$, and $\gamma_{\pm} = \epsilon'^2 \pm \sqrt{1+\epsilon'^2 \epsilon_t^{'2}}$.
The current density expression is obtained as $j_x = v_F\psi^{+}\left(\begin{matrix} \sigma_x & 0 \\ 0 & \sigma_x \end{matrix}\right)\psi $.   
The transfer matrix through any combination of a scalar and vector potential  
can now be written in terms of  transfer matrices $ \mathcal{M}_{B}(x)$ and $ \mathcal{M}_{0}(x)$ for regions with finite and zero magnetic field respectively.  This can then be used to find the transmission. The ratio of current density in region III and the incident current density in region I gives the transmission probability as a function of the angle of incidence $\phi^{\prime}=\tan^{-1}\frac{q_y}{q_x}$. 

In region I we  parametrize  $\bs{q}=(q_{x},q_{y})$  in terms of incident energy, ($\epsilon>0$) and the angle of incidence $\phi^{\prime}$ as  $(q_{x}, k_{y}-\frac{d}{2} ) = (\sqrt{ \epsilon \epsilon^{\prime}_t  + {\epsilon}^{2}}\cos \phi^{\prime}, \sqrt{ \epsilon \epsilon_t^{\prime}  + {\epsilon}^{2}}\sin \phi^{\prime})$. In region III similar parametrization can be done in terms of the angle of emergence $\theta^{\prime}$  to write $(q^{\prime}_x, k_{y}+\frac{d}{2} ) = ( \sqrt{ \epsilon \epsilon_t^{\prime}  + {\epsilon}^{2}}{\cos\theta^{\prime}}, {\sqrt{ \epsilon \epsilon_t^{\prime}  + {\epsilon}^{2}}\sin\theta^{\prime}})$. The conservation of energy and constancy of the normal component of momentum then ensure 

\beq \sin \theta^{\prime}= \frac{d}{\sqrt{\epsilon \epsilon_t^{\prime} + {\epsilon}^2}} +\sin \phi^{\prime} \nonumber \eeq

This equation explains that beyond a certain critical angle the electron beam and its cross-section (allowed $\phi$ values) is reduced with increasing barrier strength. This feature, along with chiral symmetry breaking, dictates the nature of transmission as is subsequently explained.

In figure \ref{bilayerTr} is plotted transmittance T for a magnetic barrier ((a), left column) and for a barrier $+$ well ((b), right column). In figure 1(a), the uppermost plot gives T for $B=0T$, clearly showing the region of perfect reflection around normal incidence symmetrically placed between two wings of resonant Fabry–Perot fringes. This is a generic feature of transmission through such barriers \cite{shytov}\cite{masir}. As the barrier strength increases to $B=0.2T$, a transmission region develops between these two wings due to the effect of the magnetic barrier and the resulting transmission also becomes asymmetric. For higher fields, the disappearance of perfect Klein reflection at normal incidence is seen even more clearly. We may then conclude that in BLG, though the passage from electron to hole states at normal incidence is usually forbidden due to pseudospin conservation, even a weak magnetic field lifts this restriction. For a weak magnetic field, the coupling between electron and hole states is also weak; however, for a case of resonant scattering one can expect total transmission \cite{nasgms} The total angular range of transmission however shrinks in the presence of the magnetic barrier since, beyond a certain incident angle, all electron waves suffer total internal reflection. In figure 1(b) is plotted T through a barrier $+$ well, consisting of two magnetic barriers equal in magnitude and opposite in direction such that the total flux through the region vanishes. Now, the incident and the transmitted wavevectors are parallel to each other. As a result, the net rotation of the pseudospinor due to the inhomogeneous field vanishes and Klein reflection at normal incidence is restored. The Fabry–Perot fringes bend due to asymmetric transmission at a given V.

\subsection{Conductance}

\begin{figure}
\begin{center}
\includegraphics[width=120 mm]{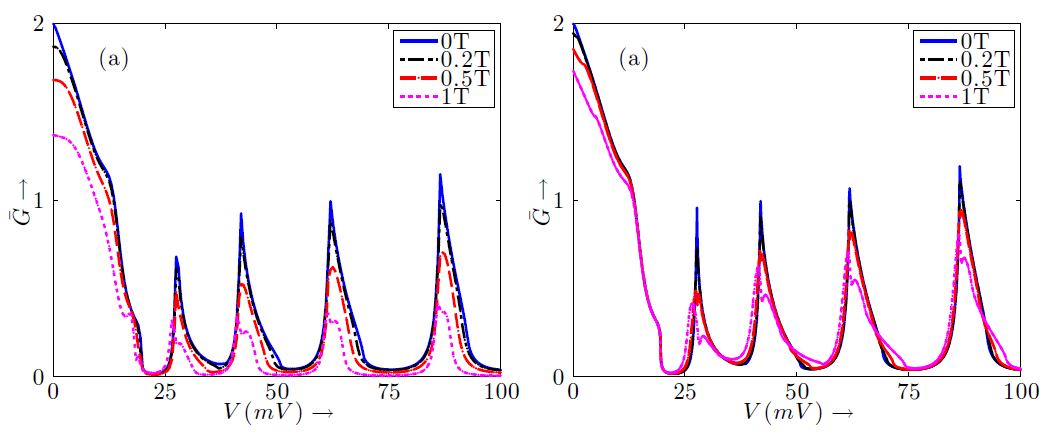}
\end{center}
\caption{${\bar{G}}$ for  different magnetic barrier strengths. (a)  Barrier,  (b) Barrier+well. $\Delta=0$,  $d=50$nm, $E_F=17$meV.
}
\label{barrierwell}
\end{figure}

We now study the effect of the above transmission on conductance at very low temperatures and for energies close to the Fermi energy in the linear transport regime. To a good approximation, the dimensionless conductance can be written as \cite{Masir1}
\beq \bar{G}=\int_{-\frac{\pi}{2}}^{\frac{\pi}{2}} d \phi T(E, \phi) \cos \phi  \eeq  
Fig.\ref{barrierwell} plots conductance $\bar{G}$ as a function voltage. This shows that the gaps between the conductance maxima and minima get reduced in the presence of a magnetic barrier. The barrier reflects electrons incident upon it beyond a critical angle  and the angular range of transmission shrinks with increasing barrier strength. As a result, the absolute value of conductance maxima comes down. Also the asymmetry, namely the way $\bar{G}$ changes to the right and left side of such maxima, reduces in the presence of magnetic barrier.
The behaviour of the  conductance is strongly dependent on transmission resonances. To explain this behaviour, therefore we shall now model these  numerically computed transmission resonances using a the Breit Wigner model \cite{breit}
\beq t(\epsilon^{\prime}, k_{y}) = \frac{\Gamma(k_{y})}{\Gamma(k_{y} )+ i(\epsilon^{\prime} - \epsilon^{\prime}_{* }(k_{y}) } \label{resonance} \eeq 
Such a form is valid near the transmission resonance. The above model has been  widely used to model the transport due to resonant tunnelling in semiconductor structure \cite{dutta}. Recently this model has been used to understand the cloaking phenomena in bilayer graphene \cite{rudner}. Here $\epsilon^{\prime}_{*}(k_{y}), \Gamma(k_{y})$ are respectively the momentum dependent resonant energy and the resonance width. 

\begin{figure}[t]
\begin{center}
\includegraphics[width=100 mm]{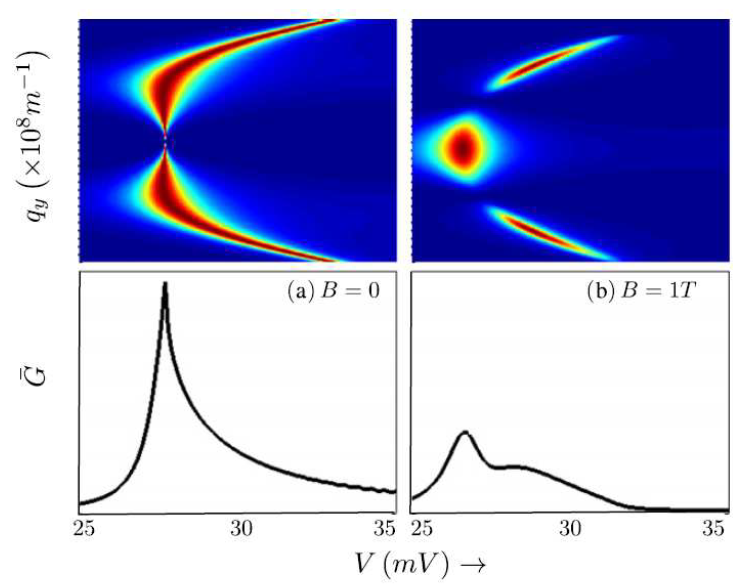}
\end{center}
\caption{Explanation for the conductance: The upper row  consists of zoomed in portion of the plot given in Fig. \ref{bilayerTr}
and the lower row consists of  the zoomed in portion of  Fig. \ref{barrierwell}.}
\label{cond2}
\end{figure}

Following the expression (\ref{dispI}) in the absence of a magnetic field, the resonance energy $\epsilon^{\prime}_{*}(k_{y})$ and the resonance width $\Gamma(k_{y})$ can be  assumed to have a quadratic dependence on the wavevector $k_{y}$ \cite{rudner}. Thus $\epsilon^{\prime}_{*}(k_{y}) \approx \epsilon^{\prime}_{*}(0) + \alpha k_{y}^2$ and $\Gamma(k_{y})= \beta k_{y}^2$. The quadratic dependence of the resonant energy and the transmission width explains why the Fabry-Perot fringes of transmission resonance depicted in the left column of Fig. \ref{cond2}  are of convex shape. The asymmetry in the form of this fringes reflects themselves in the plot of  the conductance again depicted in the left lower corner of Fig. \ref{cond2}, as one goes from the conductance peak to the left and right along the $\epsilon^{\prime}$ axis (effectively the voltage axis). In order to analyse the conductance peak, we evaluate the quantity $\delta G(\epsilon^{\prime}) =  G(\epsilon^{\prime}) - G({\epsilon^{\prime}_{*}}_{0}) $, Here $G({\epsilon^{\prime}_{*}}_{0})$ represents the conductance value at the peak, and $ G(\epsilon^{\prime})$ represents the conductance value in its neighbourhood.
Then,
\beq \delta G(\epsilon^{\prime}) = G_{0}\int\left(\left|t(\epsilon^{\prime},k_y)\right|^2- \left|t({\epsilon^{\prime}_{*}}_{0},k_y)\right|^2\right)dk_y \label{int} \nonumber \eeq

On substituting the quadratic dependencies of reseonance energy and resonance width in the integral,
\beq \delta G(\epsilon^{\prime}) = \frac{G_{0}\beta^2}{\alpha^2+\beta^2}\int_{-k_f}^{+k_f} dk_y \frac{2(\delta\epsilon^{\prime})\alpha k_y^2-(\delta\epsilon^{\prime})^2 }{\left[(\beta k_y^{2})^2 + (\delta\epsilon^{\prime}-\alpha k_y^2)^2\right]} \nonumber \eeq

which upon solving gives:
\beq \delta G(\epsilon^{\prime}) = -G_0\pi\frac{\cos[\frac{3\omega}{2}+\frac{\pi}{4}sgn(\delta\epsilon^{\prime})]}{(1+\tilde{\alpha}^2)^{3/2}}\sqrt{\frac{\left|\delta\epsilon^{\prime}\right|}{\beta}},  ~~~ \tilde{\alpha} = \alpha/\beta \eeq

Clearly $\frac{\delta G(\epsilon^{\prime})}{\delta\epsilon^{\prime}}$ shows the sqaure root sigularities at the conductance peak and the $sgn(\delta\epsilon^{\prime})$ term appearing in the expression introduces the asymmetry on the two sides of resonance energy.

In the presence of a finite magnetic barrier, the small $k_y$ behaviour of such resonances can be obtained by using a semiclassical argument. We replace $k_y$ by its covariant form, namely $k_y+2\pi\frac{Bd}{\frac{hc}{e}}$ . This yields $ \epsilon_{*}^{\prime}(k_y)\approx \epsilon_{*}^{\prime}(0)+Ad^2$ where A is a constant. With increasing strength of the magnetic barrier, $\epsilon_{*}^{\prime}(0)$ is shifted by the amount of the flux inserted in the barrier regime and this shift is independent of $k_y$. The resonance width also follows a similar behaviour. The right column of figure 3 depicts this situation. The field-induced resonance leads to a central fringe symmetric about the voltage axis, unlike the upper and lower branches, which are asymmetric. This is again seen in the conductance plot as a right-left asymmetry around the conductance peak, which is now reduced. The reduction of the conductance peak can be attributed to the shrinking of the angular range of transmission due to the magnetic barrier, whereas the rounding of the conductance peak and the consequent reduction in the left and right asymmetry around this peak is an indicator of  the reversal of Klein reflection. Since conductance is a physically measurable quantity, this provides a direct way to experimentally test our results. 
\subsection{Generalisation beyond bilayer}

\begin{figure}[t]
\begin{center}
\includegraphics[width=100 mm]{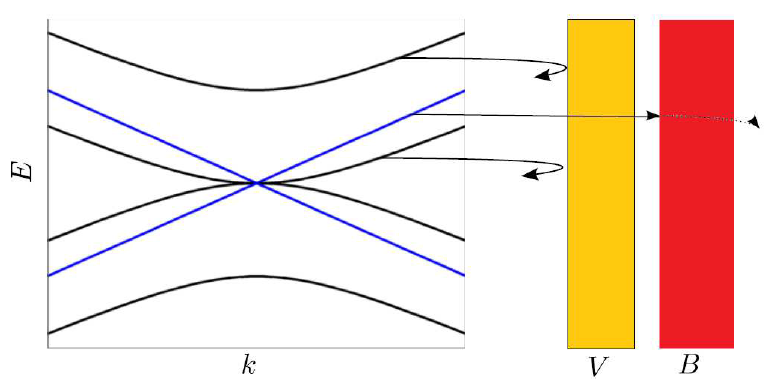}
\end{center}
\caption{Selective blocking of modes in trilayer graphene. $E$ and $k$ are in arbitrary units. The scalar potential $V$ blocks massive modes while the massless mode is passed through. The massless mode is weakly transmitted through the magnetic barrier $B$.}
\label{selectblock}
\end{figure}

The above findings also extend to multilayer graphene. For N-layer graphene with Bernal ($ABAB\cdots$)  stacking
The dispersion relation is given by \cite{min}
\beq E^{\pm}_{r,p}=t\cos (\frac{r \pi}{N+1}) \pm \sqrt{ v_F^2 |p^2|   + t^{2} \cos ^2  (\frac{r \pi}{N+1})} \nonumber \eeq 
where $ r =1,2,3,\cdots N $ corresponds to the $N$ modes of the energy spectrum, $v_{F}$ is the Fermi velocity. For odd $N$, $r = \frac{N+1}{2}$ is massless and it's energy is given by  
\beq E^{\pm}_{\frac{N+1}{2},p}= \pm v_{F} \left|p\right| \eeq
For all other $N-1$ cases 
the dispersion relation takes the form $E^2 + 2 E t \cos (\frac{r \pi}{N+1})= v^2 |p|^2 $. 
In the limit $ |E|<<{\epsilon_t}$ the low energy spectrum for the $r$-th massive mode is given 
as  \bea E_{r,p} & = & \frac{p^2}{2m_r}, \text{if} ~\epsilon_t \cos (\frac{r \pi}{N+1})  < 0  \nonumber \\
                                    & = & -\frac{p^2}{2m_r}, \text{if} ~\epsilon_t \cos (\frac{r \pi}{N+1})  < 0  \nonumber \eea 
 where  we define $m_{r}=\frac{|\epsilon_t \cos (\frac{r \pi}{N+1})|}{v_{F}^2}$ as the effective mass of the  $r$-th massive mode.
Such band structure has been plotted in fig.\ref{selectblock} for trilayer graphene. Now the transport electrons near the Fermi level will be a mixture of eigenmodes, one of which is a massless Dirac mode and the others are massive modes with different effective masses and also have a different chirality. These arguments extend to $N>3$ also.
An extension of the semiclassical argument presented earlier therefore suggests the following possibility in the case of such multilayer graphene. When such a system is exposed to a scalar potential barrier, a normally incident massless mode will undergo Klein tunnelling, whereas in the case of a massive mode there will be exponentially decaying transmission. In contrast, a magnetic barrier will be highly reflective for a massless mode but will transmit normally incident electrons for massive modes. A typical situation has been depicted in fig.\ref{selectblock}. Since the low energy band structure decomposes into one massless and $N-1$ massive modes for N odd and only N massive modes for N even \cite{min}, our results qualitatively suggest that a magnetic barrier can be used to selectively allow or partially filter out a mode.\

\begin{figure}[t]
\begin{center}
\includegraphics[width=120 mm]{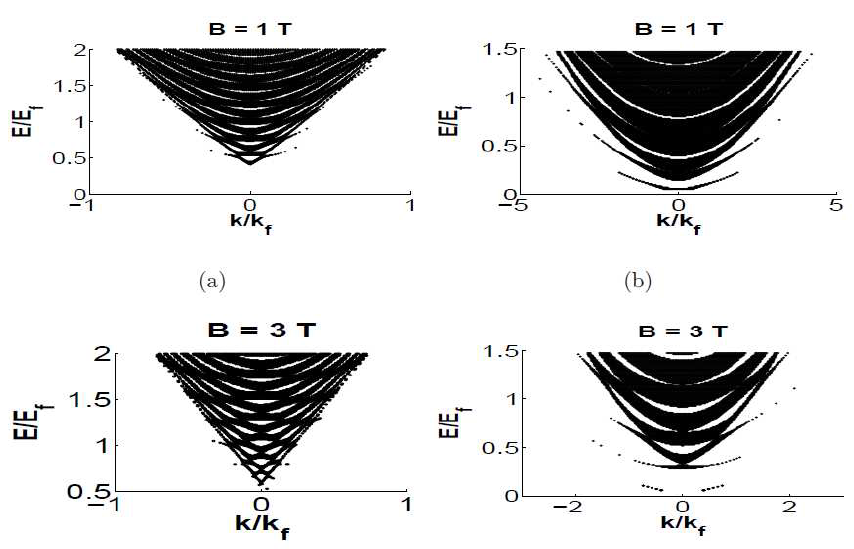}
\end{center} 
\caption{The band structure modification of monolayer and bilayer graphene in
presence of magnetic barrier. (a) and (c) plots the bandstructure of monolayer graphene in presence
of periodic arrangement of magnetic barriers with two different strengths where as (b) and (d) plots
the corresponding bandstructure of bilayer graphene.}
\label{bsemvpbilayer}
\end{figure} 

In Fig.\ref{bsemvpbilayer} we compare the modified bandstructure obtained with a periodic arrangement of DMVP in monolayer and bilayer graphene. One dimensional (1D) periodic modulation of the chemical potential or the electric field perpendicular to the layers was addressed in Ref. \cite{killi}.
 For a periodic stratified media of several DMVP barriers,  forbidden region appear in the bandstructure where the propagation of Bloch waves is not allowed. Th size of 
these forbidden regions increase with increasing strength of magnetic field. A comparison between the plots in Fig.\ref{bsemvpbilayer} (a),(c) and  (b),(d) shows very clearly
the conical band structure of monolayer graphene and parabolic bands in bilayer graphene.

This issue of anti-Klein tunnelling effect in bilayer graphene p-n-p junction and the effect of magnetic field on it, has also been addressed Ref. \cite{rudner}. Here they describe it in terms of the bound states being completely decoupled from outside continuum states, thus calling it an electronic cloaking effect, and the barrier acting as "cloak". The effect is complete at normal incidence, leading to "Klein reflection" or "anti-Klein tunneling". This is explained as follows: At normal incidence the effective equation that describes the motion of charge carriers in graphene , (as can be seen from Eq.\ref{hammat}), can be analysed in the eigen basis of $\sigma_x$: 
\beq \left(-\frac{\hbar^2}{2m}\frac{d^2}{dx^2} \pm [V-E]\right)\psi_{\pm}(x) = 0 \eeq
Clearly, the solutions for $\psi_+$ corresponds to the solutions outside the barrier. And the solution for $\psi_{-}$ (for which the equation appears with a negative sign) are bound state solutions, and corresponds to the solutions inside the barrier. Thus these two kinds of solutions are completely decoupled leading to zero transmission despite of the states available inside the barrier.

In section \ref{bilayer}, our discussion on Klein reflection in bilayer graphene was based on the analysis of Fabry-Perot like transmission resonances developed from inhomogenous magnetic barriers in the presence of voltage. Here the oscillatory transmission due to Fabry-Perot like arises due to interference of forward and backward moving propagating waves which undergo multiple reflection between two interfaces. In addition to these FP like resonances a new kind of transmission resonances have also been seen for the case of gapped bilayer graphene \cite{nandkishore}. In their recent work, R. Nandkishore et. al. proposed an entirely different approach to realize oscillatory transmission which involves only forward propagating waves and a single interface and thus is different from the FP resonances. They show that Zener transmission in a p-n junction in gapped bilayer graphene exhibits common path interference taking place
under the tunnel barrier, leading to transmission that oscillates as a function of gate-tunable bandgap.The origin of this oscillatory behaviour lies in the fact that in the presence of p-n junction, bilayer graphene employs the propagating as well as evanescent wavevector solutions. The simultaneous existence of these two- leads to common path interference which gives rise to oscillatory transmission. These oscillations in transmission manifest themselves through negative differential conductivity in the $I-V$ characteristic. 
Interesting works were also done on pseudo spinotronics in biased bilayer graphene \cite{prada, michetti} 
which we shall not discuss in this review.
Other recent developments made on these lines include the work by Campos {\it et. al.} \cite{trilayer}, where they have realized giant conductance oscillations in ballistic trilayer graphene Fabry-Perot interferometers, which result from phase coherent transport through resonant bound states beneath an electrostatic barrier.

\section{Conclusion}

A large body of work has followed since the proposal of using inhomogeneous magnetic fields in graphene and understanding the transport in an analogy with light beam propagation in optics- termed as electron optics. The similarities between polarization
states of light and ballistic charge carriers in graphene has been investigated in detail in \cite{dragomanosa, dragomanjosab29}.
It is demonstrated theoretically \cite{dragomanjosab28} that regions with different electrostatic potentials in graphene can have the same transmission probability as the transmittance of layered structures in optics illuminated with normally incident TE waves if the incidence angle of electrons in graphene and the width of the gated region are appropriately chosen.This quantitative analogy can be useful for designing optical structures that correspond to graphene-based devices. Then there are proposals of electron waveguides created by electrostatic potential or by real magnetic barriers \cite{nori,cesar}. Here the control the of the quasiparticle flux in a graphene-based waveguide was theoretically investigated. Guided modes occuring in the negative-zero-positive index metamaterial waveguide \cite{ming} and in symmetric velocity barrier \cite{jian}, have also been investigated. It is shown that magnetically induced waveguide in graphene leads to strong confinement of Dirac fermions, regardless of its edge terminations \cite{myoung}.The properties of unidirectional snake states have also been investigated in such waveguides \cite{tkghoshsnakestates}. It was shown that for certain magnetic field profile, two spatially separated counterpropagating snake states are formed, leading to conductance quantization which is insensitive to backscattering by impurities or irregularities of the magnetic field.

It may be mentioned there are many other interesting articles in the relevant field and because of space restriction 
it is almost impossible to refer to all of them which we regret. We would also like to direct the readers some excellent 
reviews on electron transport in graphene where some of the related issues  was discussed in somewhat different 
context. In ref. \cite{peresrevmodphy}  the transport properties of graphene that includes theoretical and 
experimental work was reviewed. In ref. \cite{norirev} the charge and spin transport in mesoscopic graphene structures 
was reviewd in detail. In ref. \cite{dassharma} the density and temperature-dependent charge transport in doped or gated graphene devices was reviewed.  In ref. \cite{pereirarev, fuchs} the Klein tunnelling through single and multiple potential barriers was reviewed. Properties of graphene in presence of strong homegenous magnetic field was reviewed 
in ref. \cite{goerbig}. Electronic properties of bilayer graphene was reviewed in detail in ref. \cite{mccann}. 

To summarize, We have reviewed several aspects of charge carrier transport in graphene  through inhomogenous magnetic fields viz. magnetic barriers in detail.  We reviewed experimental progress in designing such magnetic barriers at various length scales. 
Such barriers suppress Klein tunnelling, thereby achieving confinement in graphene, which can be seen through strong supression of transmission of electrons. We have shown that transport through MVP barriers can be understood in terms of propagation of light through periodic stratified media. This analogy can partially be attributed to the fact that equation describing Dirac like charge carriers in graphene and the Maxwell equations \cite{berreman} are both linear wave equations. We show that such transport properties through a singular magnetic barrier can be much better controlled by the additional application of an electrostatic voltage.  A detailed optical analogy for transport through single and multiple EMVP barriers arranged periodically was obtained highlighting optical analogues of phenomena such as TIR for positive and negative refraction and a Quantum Goos-H\"anchen shift. One significant feature of such transport in this regime of the simultaneous application of highly inhomogenous magnetic fields along with electrostatic potentials, is highly anisotropic and strongly dependent on the sign of the voltage indicating possible device applications. We discuss 
collimation, emergence of extra dirac points in this context. We also reviewed transport through such magnetic barrier
through unbiased graphene bilayer, describe reversal of Klein tunnelling, anticloaking, common path interference etc. in such 
systems.  The wide range of optical analogues to charge transport in graphene may lead to practical devices such as Bragg reflectors, resonantors, waveguides etc. We hope this detailed review will augment theoretical and experimental research in this field. 

\section{Acknowledgements}
We have benefited from discussions with many researchers including our students working on graphene in the last few years at various places. We particularly thank  G. Bhaskaran, F. M. Peeters, R. Masir, M. Barbier, L. Levitov, R. Nandkishore, Leonardo Campos, Andrea Young, P. Michetti, V. Fal'ko, E. McCann, Mandar Deshmukh, K. Sengupta, Subhasis Ghosh, Sameer Grover, Sanjay Gupta, Nupur Gupta, V. Thareja for such discussions. 
We thank A. Nogaret and IOP Publishing  for giving  us the kind permission to use one the experimental figures published in 
ref. \cite{nogaretjpcm}. This work is supported by grant SR/S2/CMP-0024/2009 from the Science and Engineering
Research Council, DST, Government of India. NA is supported by a CSIR fellowship.

\end{document}